\documentclass[aps,english,prb,floatfix,superscriptaddress,twocolumn,showpacs,amsmath,amssymb,reprint]{revtex4-1}
\usepackage[T1]{fontenc}
\usepackage[latin9]{inputenc}
\usepackage{color}
\usepackage{verbatim}
\usepackage{amsmath}
\usepackage{amssymb}
\usepackage{graphicx}

\makeatletter
\PassOptionsToPackage{caption=false}{subfig} 
\usepackage{hyperref}
\hypersetup{
breaklinks=true,
colorlinks=true,
citecolor=blue,
linkcolor=blue,
filecolor=blue,
urlcolor=blue
}
\IfFileExists{lmodern.sty}{\usepackage{lmodern}}{}
\newcommand\scalemath[2]{\scalebox{#1}{\mbox{\ensuremath{\displaystyle #2}}}}

\makeatother

\usepackage{babel}
\begin{document}

\title{A non-Abelian twist to integer quantum Hall states}

\author{Pedro L. S. Lopes}

\affiliation{Stewart Blusson Quantum Matter Institute, University of British Columbia,
Vancouver, British Columbia, Canada V6T 1Z4}

\affiliation{Institut Quantique and D\'{e}partement de Physique, Universit\'{e} de Sherbrooke,
Sherbrooke, Qu\'{e}bec, Canada J1K 2R1}

\author{Victor L. Quito}

\affiliation{Department of Physics and Astronomy, Iowa State University, Ames,
Iowa 50011, USA }

\affiliation{Department of Physics and National High Magnetic Field Laboratory,
Florida State University, Tallahassee, Florida 32306, U.S.A. }

\author{Bo Han}

\affiliation{Department of Physics and Institute for Condensed Matter Theory,
University of Illinois, 1110 W. Green St., Urbana IL 61801-3080, U.S.A.}

\author{Jeffrey C. Y. Teo}

\affiliation{Department of Physics, University of Virginia, Charlottesville, VA22904,
U.S.A.}

\begin{abstract}
At a fixed magnetic filling fraction, fractional quantum Hall states may display a plethora of interaction-induced competing phases. Effective Chern-Simons theories have  suggested the existence of multiple interaction-induced short-range entangled phases also at \emph{integer} Quantum Hall plateaus. Among these, a bosonic phase has been proposed with edge modes carrying representations based on the $E_8$ exceptional Lie algebra. Through a theoretical coupled-wire construction, we provide an explicit microscopic model for this $E_8$ Abelian quantum Hall state, at filling $\nu=16$, and discuss how it is intimately related to topological paramagnets in (3+1)D. Still using coupled wires, we partition the $E_8$ state into a pair of non-Abelian, long-range entangled states. These two states occur at filling $\nu=8$, demonstrating that \emph{even} topological order may also exist at integer Hall plateaus. These phases are bosonic, carry chiral edge theories with either $G_{2}$ or $F_{4}$ internal symmetries and host Fibonacci anyonic excitations in the bulk. This suggests that the $\nu=8$ quantum Hall plateau may provide an unexpected platform to realize decoherence-free quantum computation by anyon braiding. We also find that these topological ordered phases are related by a notion of particle-hole conjugation based on the $E_{8}$ state that exchanges the $G_{2}$ and $F_{4}$ Fibonacci states. We argue that these phases can be tracked down by their electric and thermal Hall transport satisfying a distinctive Wiedemann-Franz law $\left(\kappa_{xy}/\sigma_{xy}\right)/\left[\left(\pi^{2}k_{B}^{2}T\right)/3e^{2}\right]<1$, even at integer magnetic filling factors. 

\end{abstract}
\maketitle

\section{Introduction}\label{sec:intro}

Quantum particles are generically classified by their exchange properties, typically bosonic or fermionic. In two dimensions, however, quantum many-body interference phenomena are brought
to a new level of complexity as anyon statistics becomes a possibility~\cite{Wilczekbook,Wenbook,ChetanSimonSternFreedmanDasSarma}. In this case, the exchange of identical particles changes a system's wavefunction by a phase that may interpolate arbitrarily between the $0$ (bosonic) and $\pi$ (fermionic) limits. Fractional Quantum Hall ({\color{blue}\hypertarget{FQH}{FQH}}) fluids form the paradigmatic examples of anyonic systems. Here, topological order develops, with gapless charge and energy transporting edge-modes and with bulk excitations displaying anyonic exchange behavior~\cite{FQHE_Review}.

Due to the magnetically quenched kinetic energy of quantum Hall systems, interactions are known to drive a sensitive competition among topological phases in the \hyperlink{FQH}{FQH} regime. The $5/2$ plateau provides a standard example, where states such as the Pfaffian, anti-Pfaffian and other composite-particle pictures appear as candidates to describe the \hyperlink{FQH}{FQH} phase phenomenology~\cite{MooreRead,HLR_composite,Son_composite}. Less diversity is discussed, however, for \emph{integer} quantum Hall ({\color{blue}\hypertarget{IQH}{IQH}}) fluids. Could interactions drive topological phase transitions in Hall fluids also at integral magnetic filling fractions? A suggestively positive answer to this query was first pointed out by Kitaev~\cite{Kitaev06}. Phenomenologically quantum Hall phases conserve charge and energy. These conservations imply well-defined electric and thermal Hall transport through gapless edges, which are determined by the \emph{bulk} magnetic filling fraction $\nu$ and the \emph{edge} conformal field theory (\hyperlink{CFT}{CFT}) central charge $c$ [c.f. Eq.~\eqref{eq:kappasigma} below]. This phenomenology is well accounted for by Chern-Simons theories, from which one can also connect $\nu$ to the exchange statistics. Kitaev's finding was that for all short-range entangled ({\color{blue}\hypertarget{SRE}{SRE}}) \emph{bosonic} topological phases the chiral central charge $c$ is determined by the magnetic filling fraction $\nu$ only modulo $8$:  $c=\nu \mod 8$~\cite{Kitaev06}(c.f. also Appendix~\ref{subsec:Fibonacci-primary-fields} and the Gauss-Milgram formula discussion). The \hyperlink{IQH}{IQH} case corresponds to the limit of $c=\nu$, but phenomenology is not limited to this simplest scenario.

The developments regarding time-reversal-broken \hyperlink{SRE}{SRE} bosonic topological phases have been further explored subsequently by Lu and Vishwanath~\cite{LuVish} and Plamadeala, Mulligan and Nayak~\cite{Plama}. Both collaborations have approached this problem via a phenomenological Chern-Simons perspective. Overall, a consensus points to the existence of a bosonic phase, with edges described by a Wess-Zumino-Witten ({\color{blue}\hypertarget{WZW}{WZW}}) {\color{blue}\hypertarget{CFT}{CFT}}  based on the exceptional Lie algebra $E_{8}$ at level 1. This is the prime candidate to describe \hyperlink{SRE}{SRE} phases at integer $\nu$ that are, nevertheless, distinct from simple copies of \hyperlink{IQH}{IQH} states.  

$E_{8}$ corresponds to the largest exceptional Lie algebra, with 248 generators and representations arranged minimally in an 8-dimensional lattice~\cite{Garibaldi}. Despite this complexity, it has enjoyed attention in physical scenarios, including experimental verification in the quantum magnetism of the Ising model~\cite{ZamolodchikovE8,E8Science}. In the present context, the algebraic structure of the $E_8$ {\color{blue}\hypertarget{WZW}{WZW}} {\color{blue}\hypertarget{CFT}{CFT}} fixes the thermal Hall transport with $c=8$. Just as in Kitaev's original argument, the effective Chern-Simons approach points to phases at arbitrary values of $\nu$, all differing from $c$ by some non-zero integer multiple of 8. 

The first goal of the present work is to propose a microscopic model for this phase. To do this, we turn to an  approach based on a coupled-wire construction of quantum Hall phases, relying on a set of 1D channels forming a 2D array~\cite{KaneMukhopadhyayLubensky02}. The 2D bulk is gapped by interactions among the channels, restoring isotropy and leaving behind gapless edges. A bulk-boundary correspondence relates excitations of these gapless 1D edges to anyonic excitations in the 2D topological bulks~\cite{Kitaev06,Wen1990,MooreRead}. This method has previously succeeded in describing diverse \hyperlink{FQH}{FQH} phases~\cite{KaneMukhopadhyayLubensky02,TeoKaneCW,SirotaSahooChoTeo18} and topological superconducting phases\cite{Sahoo16,HuKane18,mongg2}. By including the features of exceptional Lie algebra embeddings, we successfully implement a coupled-wire construction for an \emph{$E_{8}$ quantum Hall state} where $c=8$ and $\nu=16$. A straightforward consequence of this discrepancy between $\nu$ and $c$ is that the $E_8$ quantum Hall state can be distinguished from the regular \hyperlink{IQH}{IQH} state ($c=\nu=16$) via the ratio between electric and thermal Hall conductivities by the  Wiedemann-Franz law~\cite{FranzWiedemann}.

While the $E_8$ state competes with the $\nu=16$ \hyperlink{IQH}{IQH} phase, it does not display a general non-Abelian topological order~\cite{Schellekens}~\footnote{technically, this follows trivially from the unimodularity of the $E_{8}$ Cartan matrix}. This prompts us to consider a more challenging scenario: could long-range topological order also develop inside an \hyperlink{IQH}{IQH} plateau? Our inclusion of exceptional Lie algebras to the coupled-wire program proves to be a fruitful tool to answer this question. We take notice of the convenient existence of a \hyperlink{CFT}{CFT} embedding of two other exceptional Lie algebras, $(G_{2})_{1}\times(F_{4})_{1}$, into $(E_{8})_{1}$~\cite{ALEXANDERBAIS1987561}.  These groups also have enjoyed recent attention in physics. Examples include the classification of particles in the standard model (see, e.g., Ref.~\onlinecite{Furey14}, and note the relationship between $G_{2}$ and the octonions algebra~\citep{HuKane18}) and, most importantly here, quantum information theory, where a connection between the $G_{2}$ and $F_{4}$ algebras and Fibonacci anyons is well-established~\cite{RowellStongWang09,mongg2,HuKane18}(see also Appendix~\ref{subsec:Fibonacci-primary-fields}). Fibonacci anyons are a holy-grail-particle in quantum information physics, offering a venue for universal (braiding-based) topological quantum computation. Using our $E_8$ construction as a parent, we build two distinct ($G_{2}$ and $F_{4}$) Fibonacci phases which compete with the \hyperlink{SRE}{SRE} \hyperlink{IQH}{IQH} phase at $\nu=8$. These Fibonacci phases are  \emph{long-range entangled}, with fractional central charges $c_{G_{2}}=14/5$ and $c_{F_{4}}=26/5$~\cite{bigyellowbook,WessZumino71,WittenWZW} and may again be probed by non-standard coefficients in the Wiedemann-Franz law. The practicality of searching for Fibonacci anyons at integer Hall plateaus should be contrasted with previous attempts at building models for Fibonacci topological order: these included the $\nu=12/5$ \hyperlink{FQH}{FQH} phase of Read and Rezayi~\citep{ReadRezayi}, a trench construction between $\nu=2/3$ \hyperlink{FQH}{FQH} and superconducting states~\citep{mongg2}, and an interacting Majorana model in a tricritical Ising coset construction~\citep{HuKane18}. While our analysis does not provide, yet, the detailed interactions in an electronic fluid picture that would lead to the Fibonacci phase, it does prove the existence of the phase at a specific and achievable $\nu=8$, bypassing \hyperlink{FQH}{FQH} phases, heterostructures, and topological superconductivity ingredients.

As a final remark, our construction shows that the $F_4$ and $G_2$ Fibonacci phases are related by an unconventional particle-hole conjugation, based on a unifying description coming from the $E_{8}$ parent phase. Fibonacci and 'anti-Fibonacci' phases have also been identified in Ref.~\onlinecite{HuKane18} and discerned by interferometric analysis. Here they can be distinguished from solely by the Wiedemann-Franz law.

\section{The $E_{8}$ quantum Hall state}\label{sec:E8state}

\begin{figure}[t!]
\includegraphics[width=0.85\columnwidth]{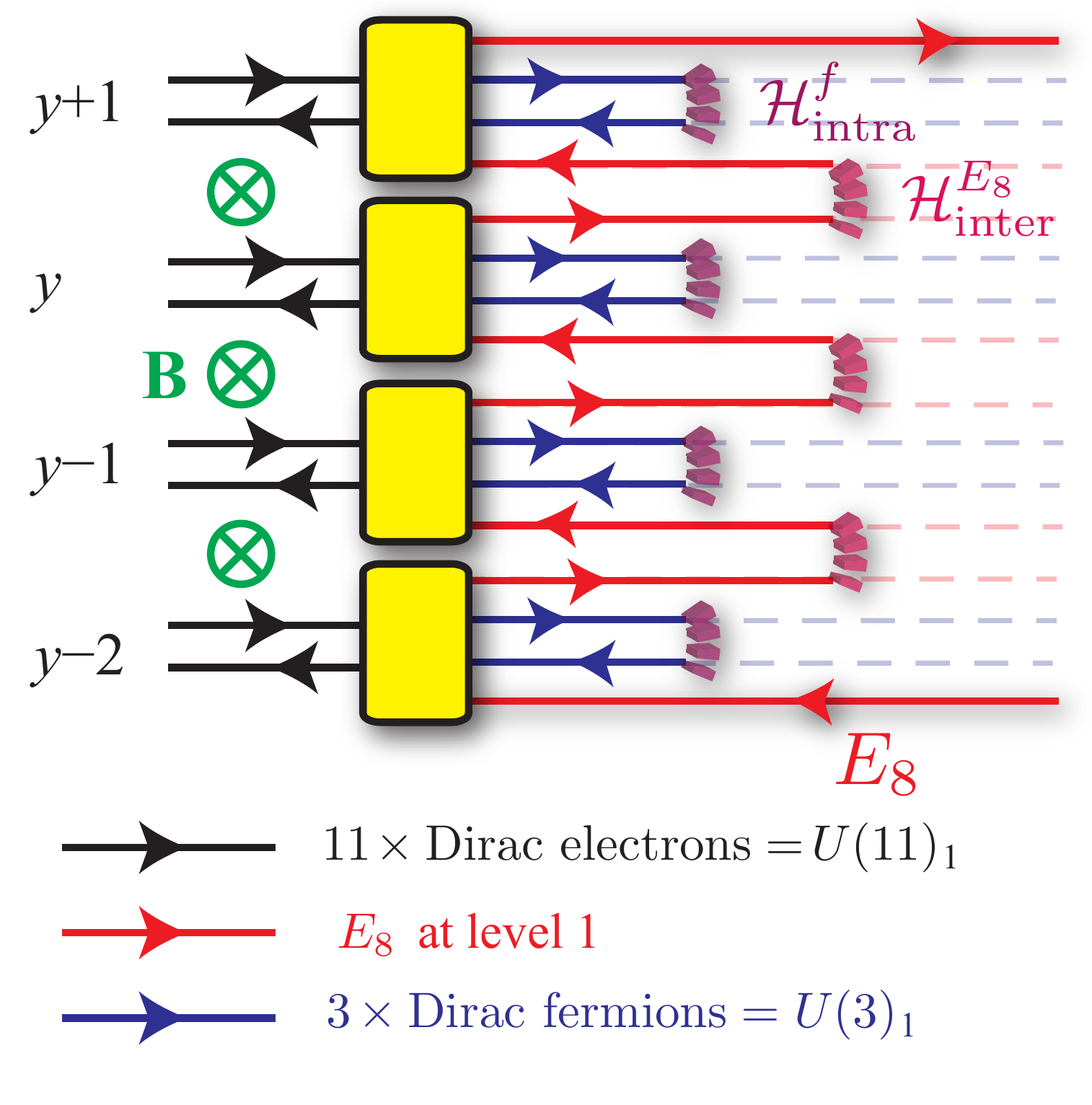}
\caption{Coupled-wire model of the $E_{8}$ quantum Hall state at filling $\nu=16$.
Black lines represent bundles with $11$ electron wires, each carrying
a counter-propagating pair of Dirac fermions, in the presence of a
magnetic flux (green). Yellow boxes represent an unimodular basis
transformation $U$ ($\det(U)=1$) restructuring $U(11)_{1}\to U(3)_{1}\times(E_{8})_{1}$.
The spectator fermionic $U(3)_{1}$ triplets and the bosonic $(E_{8})_{1}$
are coupled through intra-bundle and inter-bundle backscatteings $\mathcal{H}_{\mathrm{intra}}$
and $\mathcal{H}_{\mathrm{inter}}$ defined in \eqref{eq:IntraU1}
and \eqref{eq:InterE8}. The 2D bulk is fully gapped leaving just
the chiral $(E_{8})_{1}$ modes at the edges.}
\label{fig:Coupled-wires-construction} 
\end{figure}

Our construction begins with an array of electron wires in bundles (Fig.~\ref{fig:Coupled-wires-construction} black lines) with vertical positions $\mathsf{y}=dy$, $d$ being their displacement and $y$ an integer label. Each bundle contains $N$ wires carrying, at the Fermi level, left ($L$) and right ($R$) moving fermions whose annihilation operators admit a bosonized representation
\begin{align}
c_{ya}^{\sigma}\left(\mathsf{x}\right)\sim\exp\left[i\left(\Phi_{ya}^{\sigma}\left(\mathsf{x}\right)+k_{ya}^{\sigma}\mathsf{x}\right)\right],\label{elecboson}
\end{align}
forming a $U(N)_{1}$ \hyperlink{WZW}{WZW} theory. Here, $a=1,\ldots,N$ labels the
wires, $\mathsf{x}$ is the coordinate along them, $\sigma=R,L=+,-$
is the propagation direction and $k_{ya}^{\sigma}$ is the Fermi momentum
of each channel. The bosonic variables obey the commutation relations
\begin{align}
\left[\partial_{\mathsf{x}}\Phi_{ya}^{\sigma}\left(\mathsf{x}\right),\Phi_{y'a'}^{\sigma'}\left(\mathsf{x}'\right)\right] & =2\pi i\sigma\delta^{\sigma\sigma'}\delta_{aa'}\delta_{yy'}\delta\left(\mathsf{x}-\mathsf{x}'\right).\label{ETCR1}
\end{align}

To couple the fermions of different bundles and introduce a finite
excitation energy gap, while leaving behind gapless chiral $(E_{8})_{1}$
edges, two ingredients are necessary: {\color{blue}\hypertarget{(i)}{(i)}} a basis transformation that
extracts the $(E_{8})_{1}$ degrees of freedom from $U(N)_{1}$ (Fig.~\ref{fig:Coupled-wires-construction}
yellow boxes) and {\color{blue}\hypertarget{(ii)}{(ii)}} backscattering interactions between $L$- and
$R$-movers of different bundles to gap out all low energy channels throughout the bulk (Fig.~\ref{fig:Coupled-wires-construction} dashed arcs).

\begin{figure}[t!]
\centering\includegraphics[width=1.0\columnwidth]{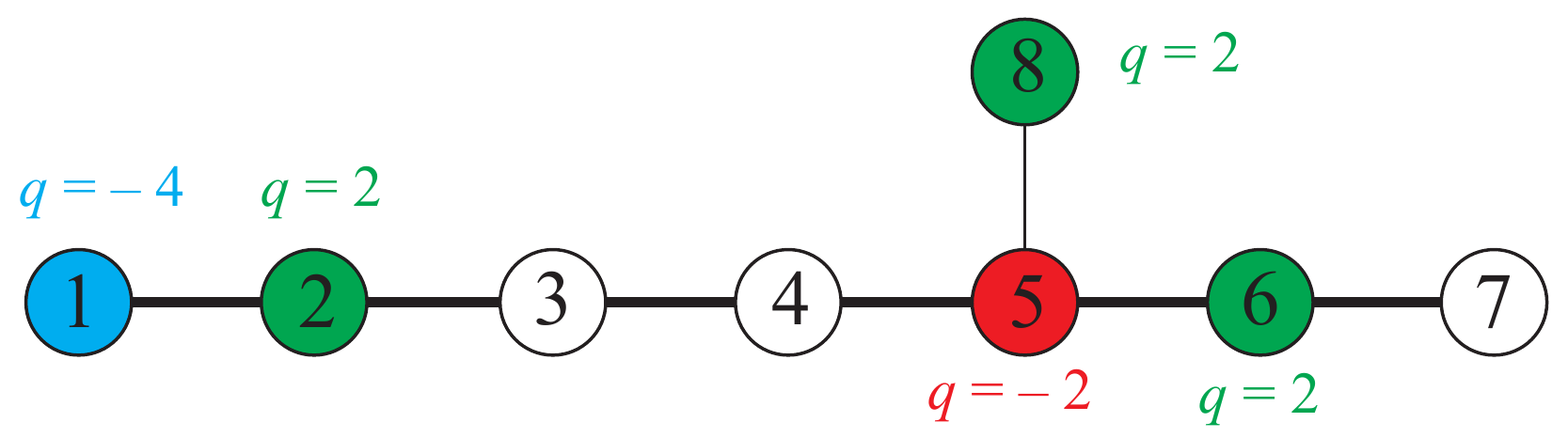}
\caption{The Dynkin diagram of $E_{8}$ and the charge assignment $q$ (in
units of $e$) of the simple roots $(E_{E_{8}})_{y\boldsymbol{\alpha}_{I}}^{\sigma}$,
for $I=1,\ldots,8$. Uncolored entries are electrically neutral.}
\label{fig:Local-bosonic-embedding} 
\end{figure}

For ingredient \hyperlink{i}{(i)}, the bosonization approach provides a convenient solution. Out of the 240 $E_8$ off-diagonal current operators, it suffices to generate the 8 simple roots, basis of the $E_{8}$ root lattice. These assume, under bosonization,
the general form~\cite{bigyellowbook} 
\begin{align}
\left[E_{E_{8}}\right]_{y\boldsymbol{\alpha}_{I}}^{\sigma}\sim\exp\left[i\left(\tilde{\Phi}_{yI}^{\sigma}\left(\mathsf{x}\right)+\tilde{k}_{yI}^{\sigma}\mathsf{x}\right)\right],\,I=1,...,8.\label{eq:E8sroots}
\end{align}
Here $\boldsymbol{\alpha}_{I}$ is a simple root vector of $E_{8}$
so that 
\begin{align}
\left[\partial_{\mathsf{x}}\tilde{\Phi}_{yI}^{\sigma}(\mathsf{x}),\tilde{\Phi}_{y'I'}^{\sigma'}(\mathsf{x}')\right]=2\pi i\sigma\delta^{\sigma\sigma'}K_{II'}^{E_{8}}\delta_{yy'}\delta(\mathsf{x}-\mathsf{x}'),\label{ETCR2}
\end{align}
and $K_{II'}^{E_{8}}=\boldsymbol{\alpha}_{I}\cdot\boldsymbol{\alpha}_{I'}$
is the $E_{8}$ Cartan matrix
\begin{equation}
K^{E_{8}}=\left(\begin{array}{cccccccc}
2 & -1\\
-1 & 2 & -1\\
 & -1 & 2 & -1\\
 &  & -1 & 2 & -1\\
 &  &  & -1 & 2 & -1 &  & -1\\
 &  &  &  & -1 & 2 & -1\\
 &  &  &  &  & -1 & 2\\
 &  &  &  & -1 &  &  & 2
\end{array}\right).
\end{equation}
The challenge now is to represent the $E_{8}$ roots as products of electron operators, so that their bosonized variables are related to the electronic ones by an integer-valued transformation $\tilde{\Phi}_{yI}^{\sigma}=U_{Ia}^{\sigma\sigma'}\Phi_{ya}^{\sigma'}$. As a consistency condition from \eqref{ETCR1} and \eqref{ETCR2}, $\sigma''U_{Ia}^{\sigma\sigma''}U_{I'a}^{\sigma'\sigma''}=\sigma\delta^{\sigma\sigma'}K_{II'}^{E_{8}}$. From \eqref{elecboson}, the $E_8$ roots momenta and charges are related to the fermionic ones,
\begin{equation}
\tilde{k}_{yI}^{\sigma}=U_{Ia}^{\sigma\sigma'}k_{ya}^{\sigma'}
\end{equation}
and 
\begin{equation}
\tilde{q}_{I}^{\sigma}=U_{Ia}^{\sigma\sigma'}q_{a}^{\sigma'},\label{eq:charges} 
\end{equation} respectively. Such a basis transformation exists, but is not unique, and requires, in particular, $N>8$ wires. To fix a solution, we demand the extra modes to correspond to a trivial fermionic sector. This way, a possible construction contains $N=11$ wires, decomposing into a $E_8$ and three $U(1)$ sectors~\footnote{In fact, a solution exists for $N=9$ wires also, where the $E_{8}$
quantum Hall phase develops at filling fraction $\nu=32$, higher
than our present solution. Also, the Dynkin labels 4, 5, 6 and 8 in this construction are neutral, leaving an SO(8) subsector with trivial $\mathsf{x}$-momenta. A main consequence is that the embedding of $G_{2}$ currents also carry trivial momenta, and Fibonacci phases can never be stabilized.}.
In practice, we write
\begin{align}
U=\begin{pmatrix}U^{++} & U^{+-}\\
U^{-+} & U^{--}
\end{pmatrix}\label{Umatrix}
\end{align}
 as unimodular matrix, decomposing $U\eta U^{T}=K^{E_{8}}\oplus\openone_{3}\oplus(-K^{E_{8}})\oplus(-\openone_{3})$,
where $\eta^{\sigma\sigma'}=\sigma\delta^{\sigma\sigma'}$. For our particular construction,

\begin{align}
 & (U^{++}|U^{+-})=(U^{--}|U^{-+})=\label{eq:Mmat}\\
 & \left(\scalemath{0.62}{\begin{array}{ccccccccccc|ccccccccccc}
-1 & -1 & -1 &  &  &  &  &  &  &  &  &  &  &  &  &  &  &  &  &  &  & -1\\
 &  & 1 & 1 &  &  &  &  &  &  & \\
 &  &  & -1 & 1 &  &  &  &  &  & \\
 &  &  &  & -1 & 1 &  &  &  &  & \\
 &  &  &  &  & -1 & -1 &  &  &  & \\
 &  &  &  &  &  & 1 & 1 &  &  & \\
 &  &  &  &  &  &  & -1 & 1 &  & \\
 &  & -1 & 1 & 1 & 1 &  &  &  &  &  &  &  &  &  &  &  &  &  &  & 1 & -1\\
 &  &  &  &  &  &  &  &  & 1 & 1 & 1\\
 &  &  &  &  &  &  &  &  & 3 & -5 & -2 & -1 & -2 & 2 & 2 & 2 & -2 & 2 & 2\\
2 &  & 1 & -1 & -1 & -1 & 1 & -1 & -1 &  &  &  &  &  &  &  &  &  &  &  & -1 & 3
\end{array}}\right)\nonumber 
\end{align}
where the rows and columns of $U^{\sigma\sigma'}$ are respectively
labeled by $I,a=1,\ldots,11$. Rows $I=1$ to $8$ associate to the
simple roots of $E_{8}$. Substituting the unit electric charge $q^{\sigma}_{a}=1$ for all electronic channels in Eq.~\eqref{eq:charges}, we find the electric charge assignments 
\begin{equation}
\tilde{{\bf q}}^\sigma=(-4,2,0,0,-2,2,0,2) \label{eq:chargevec}
\end{equation}
carried by the eight $E_8$ simple roots of each chiral sector; these may be conveniently organized in the corresponding Dynkin diagram as in Fig~\ref{fig:Local-bosonic-embedding}. Rows 9 to 11 correspond to  Dirac fermions (spin $|h|=1/2$) $f_{yn}^{\sigma}\sim\exp\left[iU_{I=8+n,a}^{\sigma\sigma'}(\Phi_{ya}^{\sigma'}+k_{ya}^{\sigma'}\mathsf{x})\right]$, for $n=1,2,3$, that generate $U(3)_{1}$. They are also integral products of the original electrons and carry odd electric charges $(\tilde{q}_{n=1,..,3})=(3,1,1)$, calculated using the same steps that lead to Eq.~\eqref{eq:chargevec}. 

Returning now to ingredient \hyperlink{ii}{(ii)}, electron backscattering interactions generally require momentum commensurability to stabilize oscillatory factors~\citep{Fradkinbook}. To tune these phases, and break time-reversal as necessary in a quantum Hall fluid, we introduce a magnetic field perpendicular to the system (Fig.~\ref{fig:Coupled-wires-construction}
green crosses). The Fermi momenta of the electron channels become
spatially dependent as 
\begin{align}
k_{ya}^{\sigma}=\frac{eB}{\hbar c}\mathsf{y}+\sigma k_{F,a}.\label{Fermimomenta}
\end{align}
We choose the Lorenz gauge where $A_{x}=-B\mathsf{y}$ and label the bare bare Fermi momenta in the absence of field as $k_{F,a}$. The associated magnetic
filling fraction can be expressed as 
\begin{equation}
\nu=
\frac{\frac{1}{2\pi}\sum_{a}2k_{F,a}}{Bd/\phi_{0}}=\frac{\hbar c}{eBd}\sum_{a}2k_{F,a},
\end{equation}
where $\phi_{0}=hc/e$ is the magnetic flux quantum.

At this point, we introduce the wire-coupling interactions 
\begin{align}
\mathcal{H}_{\mathrm{intra}}^{y,f} & =u_{\mathrm{intra}}\sum_{n=1}^{3}{f_{yn}^{R}}^{\dagger}f_{yn}^{L}+h.c.,\label{eq:IntraU1}\\
\mathcal{H}_{\mathrm{inter}}^{y+1/2,E_{8}} & =u_{\mathrm{inter}}\sum_{I=1}^{8}{\left[E_{E_{8}}\right]_{y,\boldsymbol{\alpha}_{I}}^{R}}^{\dagger}\left[E_{E_{8}}\right]_{y+1,\boldsymbol{\alpha}_{I}}^{L}+H.c..\label{eq:InterE8}
\end{align}
From \eqref{eq:E8sroots}, and the corresponding bosonization of $f_{yn}^{\sigma}$, Eqs.~\eqref{eq:IntraU1} and \eqref{eq:InterE8}
carry momentum-dependent oscillating factors $e^{ik\mathsf{x}}$
which average to zero in the thermodynamic limit. Demanding the absence
of these oscillations, i.e.~requiring the backscattering interactions
to conserve momentum, leads to the set or equations
\begin{align}
\left(U_{I,a}^{++}-U_{I,a}^{+-}\right)\left(k_{ya}^{R}-k_{y+1a}^{L}\right)=0 \,\,\, I=1,...,8 \nonumber\\ 
\left(U_{I,a}^{++}-U_{I,a}^{+-}\right)\left(k_{ya}^{R}-k_{ya}^{L}\right)=0 \,\,\,\,\, I=9,10,11, \label{eq:E8momenta_fix}
\end{align}
whose solution fixes the ratios between the bare $k_{F,a}$ uniquely and, most remarkably, also fixes uniquely $\nu=16$. The values of these momenta are listed in Appendix~\ref{subsec:App_momenta}.

It is worth to note that the charge vector of Eq.~\eqref{eq:chargevec} allows a consistency checking of the $\nu=16$ magnetic filling fraction.  According to the effective Chern-Simons field theory approach, the filling fraction is uniquely determined by the K-matrix and quasi-particle charges by
 
\begin{equation}
\nu=\tilde{{\bf q}}^{T}(K^{E_{8}})^{-1}\tilde{{\bf q}},
\end{equation}
where a single chiral sector is used (we omit the label), and where $K^{E_{8}}$ is the $E_{8}$ Cartan matrix. The filling fraction $\nu=16$ comes from this equation and the momentum commensurability condition has, again, a unique solution (up to a single free Fermi-momentum parameter $k_{F}$). 

Under the conditions above, and in a periodic geometry with $N_{l}$ bundles, the intra- and inter-bundle backscattering Hamiltonians introduce $11\times N_{l}$ independent sine-Gordon terms satisfying the Haldane's nullity condition~\cite{Haldane95}. These interactions are generically irrelevant in the renormalization group sense, although this may change in the presence of forward scattering and velocity terms. At strong coupling, however, they lead to a finite energy excitation gap in the coupled-wire model. Also, these interactions are favored over several other simpler interaction terms due to the momentum commensurability conditions.

The $E_{8}$ quantum Hall phase carries distinctive phenomenology. Opening the periodic boundary conditions
leaves behind, at low energies, eight chiral $E_{8}$ boundary modes along the top and bottom edges,
as illustrated in Fig.~\ref{fig:Coupled-wires-construction}. As consequence of the discrepancy between the magnetic filling factor and the number of $E_8$ edge modes, we predict an unconventional Wiedemann-Franz law~\cite{FranzWiedemann} for the $E_8$ quantum Hall phase: a general set of gapless edge modes, as in regular \hyperlink{IQH}{IQH} states, carries the differential thermal and electric conductances (or, equivalently, Hall conductances)~\cite{Halperin82,WenchiralLL90,KaneFisher97,Cappelli01,Kitaev06}
\begin{align}
\kappa_{xy}=c\frac{\pi^{2}k_{B}^{2}}{3h}T,\quad\sigma_{xy}=\nu\frac{e^{2}}{h}, \label{eq:kappasigma}
\end{align}
where $e$ is the electric charge, $h$ is Planck's
constant, $k_{B}$ is Boltzmann constant, $c$ is the
chiral central charge and $T$ is the temperature. For a standard
\hyperlink{IQH}{IQH} state, $c=\nu$ identical to the number of chiral Dirac
electron edge channels. A deviation away from $c/\nu=1$ indicates
the onset of a strongly-correlated many-body phase. Here, the $E_{8}$
quantum Hall phase carries 8 chiral edge bosons and therefore $c_{E_8}=8$,
while $\nu=16$ is necessary to stabilize the phase. This leads to
a modified Wiedemann-Franz law, where $c_{E_8}/\nu=1/2$.

We note in passing that the $E_{8}$ state is topologically related
to a thin slab of a 3D $e_{f}m_{f}$ topological paramagnet with time-reversal symmetry-breaking
top and bottom surfaces~\cite{VishwanathSenthil12,WangPotterSenthil13}.
Like a topological insulator, hosting a 1D chiral Dirac channel with $(c,\nu)=\pm(1,1)$ along
a magnetic surface domain wall, the $e_{f}m_{f}$ topological paramagnet
supports a neutral chiral $E_{8}$ interface with $(c,\nu)=\pm(8,0)$
between adjacent time-reversal breaking surface domains with opposite
magnetic orientations~\cite{HasanKane10,QiZhangreview11,HasanMoore11,RMP}. Comparing $(c,\nu)=(8,16)=(16,16)-(8,0)$,
the charged edge modes of the $E_{8}$ quantum Hall state are therefore
equivalent to the neutral $E_{8}$ topological paramagnet surface
interface up to 16 chiral Dirac channels, which exists on the edge
of the conventional $\nu=16$ \hyperlink{IQH}{IQH} state. In fact, the matrices $K^{E_{8}}$
and $\openone_{16}\oplus(-K^{E_{8}})$ are related by a charge preserving
stable equivalence~\cite{cano2013bulk}. Finally, the unimodularity
of the $E_{8}$ lattice entails that all primary fields of the edge
$E_{8}$ \hyperlink{CFT}{CFT} are integral products of the simple roots \eqref{eq:E8sroots},
which are even products of electron operators. Hence, ignoring any
edge reconstruction, the edge modes of the $E_{8}$ state support {\em only} 
evenly charged bosonic gapless excitations.

\section{The Fibonacci states}\label{sec:Fib_states}

The $E_{8}$ state construction above serves as a stepping stone for building coupled-wire models of other phases based on exceptional Lie algebras. Here, we focus on demonstrating the existence of phases carrying $\left(G_{2}\right)_{1}$ or $\left(F_{4}\right)_{1}$ \hyperlink{WZW}{WZW} \hyperlink{CFT}{CFT}s at the edges, again at integer magnetic filling fractions. Remarkably, these phases correspond to Fibonacci topological order (c.f. Appendix \ref{subsec:Fibonacci-primary-fields}). To build these models, we proceed with a conformal embedding of $G_{2}\times F_{4}$ into $E_{8}$. The existence of such embedding is signaled by the relationship among central charges $c_{E_{8}}=8=14/5+26/5=c_{G_{2}}+c_{F_{4}}$; a rigorous proof of its existence is possible can be found in Ref.~\onlinecite{ALEXANDERBAIS1987561}. Conversely, the $G_{2}$ or $F_{4}$ Fibonacci phases can be thought as arising from a partition or fractionalization of the $E_8$ parent state. In what follows, we start by displaying an explicit construction of the conformal embedding. We then follow with the coupled-wire construction and finish with an analysis of a particle-hole relation between the two Fibonacci states.

\subsection{A \texorpdfstring{$G_{2}\times F_{4}$}{G2,F4} conformal embedding
into \texorpdfstring{$E_{8}$}{E8}}

The conformal embedding is carried out by an explicit choice of the generators of $F_{4}$ and $G_{2}$, denoted by $[E_{F_{4}}]_{y,\boldsymbol{\alpha}}^{\sigma}$ and $[E_{G_{2}}]_{y,\boldsymbol{\alpha}}^{\sigma}$, where $\boldsymbol{\alpha}$ are vectors in the $F_{4}$ or $G_{2}$ root lattices $\Delta_{F_{4}}$ or $\Delta_{G_{2}}$, respectively. This process is not unique. Intuitively, it can be understood as follows: algebraically, $G_{2}\subseteq SO\left(7\right)\subseteq SO\left(16\right)\subseteq E_{8}$, i.e. $G_{2}$ is `slightly smaller` and fits inside $SO\left(7\right)$. Conversely, $SO(9)\subseteq F_{4}\subseteq E_{8}$. Altogether, one has $SO(7)\times SO(9)\subseteq SO(16)\subseteq E_{8}$. The path to follow becomes then salient: first we refermionize the $E_{8}$ generators of Eq.~(\ref{eq:E8sroots}) into bilinear products of 8 non-local Dirac fermions $d_I$. Decomposing these into Majorana components as $d_{I}=(\psi_{2I-1}+i\psi_{2I})/\sqrt{2}$, $I=1,...,8$, we obtain a representation of $SO(16)_1$. These are the degrees of freedom that we need and we can then easily accommodate a specific choice splitting $SO(16)_1=SO(7)_1\times SO(9)_1$, and then embeding $G_{2}$ into $SO(7)$ and extending $SO(9)$ into $F_{4}$. Let us follow this step-by-step.

\underline{\texorpdfstring{From $E_{8}$ to $SO\left(16\right)$}{E8,SO16}} - The $E_8$ current algebra is fixed by its 8 mutually commuting Cartan operators and its $E_8$ 240-dimensional root  lattice denoted by $\Delta_{E_{8}}$. The roots act as raising and lowering operators of the ``spin'' (weights) eigenvalues.
Let us relate the bosonized description of the $E_{8}$ \hyperlink{WZW}{WZW}
current algebra at level 1 based on the 8 aforementioned simple roots
in (\ref{eq:E8sroots}) to the desired SO(16) embedding.

We begin by {\em fermionizing} the 8 simple roots operators. This
expresses each $E_{8}$ root as either a pair or a half-integral combination
of a set of 8 {\em non-local} Dirac fermions $d_{yI}^{\sigma}\sim\exp\left[i(\phi_{yI}^{\sigma}(\mathsf{x})+k_{yI}^{\sigma}\mathsf{x})\right]$.
The bosonized variables and momenta are related to those of the 8 simple
roots by 
\begin{align}
\tilde{\Phi}_{yI}^{\sigma}=R_{I}^{I'}\phi_{yI'}^{\sigma},\quad\tilde{k}_{yI}^{\sigma}=R_{I}^{I'}k_{yI'}^{\sigma},\label{Rtrans}
\end{align}
where the $8\times8$ $R$ matrix is
\begin{align}
R=\left(\begin{smallmatrix}1 & -1\\
 & 1 & -1\\
 &  & 1 & -1\\
 &  &  & 1 & -1\\
 &  &  &  & 1 & -1\\
 &  &  &  &  & 1 & 1\\
-\frac{1}{2} & -\frac{1}{2} & -\frac{1}{2} & -\frac{1}{2} & -\frac{1}{2} & -\frac{1}{2} & -\frac{1}{2} & -\frac{1}{2}\\
 &  &  &  &  & 1 & -1
\end{smallmatrix}\right)\label{Rmatrix}
\end{align}
The lines of the $R$ matrix form a set of primitive basis vectors that are commonly adopted to generate the $E_{8}$ root lattice in $\mathbb{R}^{8}$.

The $R$ matrix decomposes the Cartan matrix $K^{E_{8}}$ of $E_{8}$
as $K^{E_{8}}=RR^{T}$. Consequently, under the transformation (\ref{Rtrans}),
the equal-time commutation relation (\ref{ETCR2}) becomes
\begin{equation}
\left[\partial_{\mathsf{x}}\phi_{yI}^{\sigma}(\mathsf{x}),\phi_{y'I'}^{\sigma'}(\mathsf{x}')\right]=2\pi i\sigma\delta^{\sigma\sigma'}\delta_{II'}\delta_{yy'}\delta(\mathsf{x}-\mathsf{x}'). 
\end{equation}
This ensures the vertex operators $d_{yI}^{\sigma}\sim\exp\left[i(\phi_{yI}^{\sigma}(\mathsf{x})+k_{yI}^{\sigma}\mathsf{x})\right]$ to represent complex Dirac fermions. As we argue next, these fermions do not associate to natural excitations in the bulk or the edge of
the quantum Hall states. Inverting the matrix (\ref{Rmatrix}) and multiplying by the original unimodular transformation~\eqref{Umatrix}, one sees that all $\phi_{yI}^{\sigma}$ expressed in terms of the original electronic
bosonized variables $\Phi_{ya}^{\sigma}$ involve half-integral coefficients. This non-locality is also revealed by their even charge assignments $q=0,\pm2$. The pair creation of such non-local Dirac fermions requires a linearly divergent energy in the coupled-wire model and, as a result, these fermions do not arise as deconfined bulk excitations or gapless edge primary fields. They should only be treated as artificial fields introduced to describe the \hyperlink{WZW}{WZW} current algebra. 

By decomposing the 8 Dirac fermions into 16 Majorana fermions as $d_{I}=(\psi_{2I-1}+i\psi_{2I})/\sqrt{2}$ (henceforth, where it leads to no confusion, we are suppressing the $\sigma,y$ indices for conciseness.)
 The $E_{8}$ \hyperlink{WZW}{WZW} current algebra can be related to an $SO(16)_{1}$
\hyperlink{WZW}{WZW} current algebra. In terms of root systems, $\Delta_{E_{8}}$ is
shown to be an extension of $\Delta_{SO(16)}$, as follows. The root
lattice of $SO(16)_{1}$, $\Delta_{SO(16)}$, contains $2^{2}\times C_{2}^{8}=112$
elements, with $C_{n}^{k}$ being the binomial coefficient. The elements
are given by bosonic spin 1 fermion pairs $d_{I}^{\pm}d_{I'}^{\pm}\sim e^{i(\pm\phi_{I}\pm\phi_{I'})}$,
where $1\leq I<I'\leq8$.  Besides the root system of
$SO(16)_{1}$, to generate the root system of $\Delta_{E_{8}}$we
include the $128=2^{7}$ even SO(16) spinors. The even spinors can
be represented by bosonic spin 1 half-integral combinations $d_{I}^{\epsilon^{I}/2}\sim e^{i\epsilon^{I}\phi_{I}/2}$,
where $\epsilon^{I}=\pm1$ and $\prod_{I=1}^{8}\epsilon^{I}=+1$.
By combining with the even spinors of the root lattice of $SO(16)$, the
$112+128=240$ roots of $E_{8}$ can be represented by the vertex
operators 
\begin{align}
[E_{E_{8}}]_{y\boldsymbol{\alpha}}^{\sigma} & \sim  \exp\left[i\alpha^{I}(\phi_{yI}^{\sigma}(\mathsf{x})+k_{yI}^{\sigma}\mathsf{x})\right] \nonumber \\ 
& = \exp\left[i\alpha^{I}(R^{-1})_{I}^{I'}U_{I'a}^{\sigma\sigma'}(\Phi_{ya}^{\sigma'}(\mathsf{x})+k_{ya}^{\sigma'}\mathsf{x})\right],\label{E8rootrep}
\end{align}
where the root vectors $\boldsymbol{\alpha}=(\alpha^{1},\ldots,\alpha^{8})$
are
\begin{align}
\scalemath{0.88}{\Delta_{E_{8}} =\left\{ \boldsymbol{\alpha}\in\mathbb{Z}^{8}:|\boldsymbol{\alpha}|^{2}=2\right\} \cup\left\{ \boldsymbol{\alpha}=\frac{\boldsymbol{\epsilon}}{2}:\epsilon^{I}=\pm1,\prod_{I=1}^{8}\epsilon^{I}=1\right\}} .\label{fullE8roots}
\end{align}
Each root vector $\boldsymbol{\alpha}$ can be expressed as a linear
combination $\alpha^{J}=a^{I}R_{I}^{J}$, with the $R$ matrix given
in (\ref{Rmatrix}) and $a^{I}$ integer coefficients, which are the
entries of the root vectors in the Chevalley basis. This integer combination ensures
that every $E_{8}$ root operator in (\ref{fullE8roots}) is an integral
combination of local electrons (\ref{elecboson}). Since each of these
vertex operators is a spin-1 boson, it must be an even product of
electron operators and therefore must carry even electric charge.

The fermionization of the $E_{8}$ presented above allows us to represent
all the $E_{8}$ roots using a vertex operator $[E_{E_{8}}]_{y\boldsymbol{\alpha}}^{\sigma}\sim\exp\left[i\alpha^{I}(\phi_{yI}^{\sigma}+k_{yI}^{\sigma}\mathsf{x})\right]$
(see (\ref{E8rootrep})), where $d_{yI}^{\sigma}\sim\exp\left[i(\phi_{yI}^{\sigma}+k_{yI}^{\sigma}\mathsf{x})\right]$
are 8 non-local Dirac fermions and $\boldsymbol{\alpha}$ are Cartan-Weyl
root vectors in $\Delta_{E_{8}}$ (recall (\ref{Rtrans}) and (\ref{fullE8roots})).
To complete the algebra structure, the 8 Cartan generators of $E_{8}$,
which are identical to the Cartan generators of $SO(16)$, are given
by the number density operators $[H_{E_{8}}]_{yI}^{\sigma}\sim i\partial\phi_{yI}^{\sigma}\sim(d_{yI}^{\sigma})^{\dagger}d_{yI}^{\sigma}$.
This also allows an explicit conformal embedding of the $G_{2}$ and
$F_{4}$ \hyperlink{WZW}{WZW} \hyperlink{CFT}{CFT}s in the $E_{8}$ theory at level 1.

\underline{From \texorpdfstring{$SO\left(16\right)$}{Lg} to \texorpdfstring{$G_{2}\times F_{4}$}{Lg}} -  We are ready to analyze the $G_{2}$ and $F_{4}$ constructions. First, since $G_{2}\subseteq SO\left(7\right)$, the $(G_{2})_{1}$ current operators have free field representations using $\psi_{1},...,\psi_{7}$, which generate $SO(7)_{1}$. Second, $SO(9)\subseteq F_{4}$. The work is a little more involved in this case: the root system of $F_{4}$ composes of (i) 24 (long) roots, (ii) 8 vectors, and (iii) 16 (even and odd) spinors of $SO(8)$, all of which may act on $\psi_{9},...,\psi_{16}$. As we will see below, accompanying the $SO(8)$ vectors with the remaining Majorana $\psi_{8}$ in $SO(9)$ and  with two special emergent fermions, we are able to to embed the $F_{4}$ currents in $E_{8}$ in a way that is fully decoupled from $G_{2}$. To abridge, $G_{2}$ is a 'bit smaller' than $SO(7)$ while $F_{4}$ is a 'bit bigger' than $SO(9)$, and the two \hyperlink{WZW}{WZW} algebras at level 1 completely decomposes $(E_{8})_{1}$.

To construct the embedding explicitly, we start by representing the
SO(7) Kac-Moody currents with Majorana fermions as $J_{SO\left(7\right)}^{a}=-i:\psi_{i}\Lambda_{ij}^{a}\psi_{j}:/2$,
where $\Lambda^{a}$ are generators of the SO(7) Lie algebra. We introduce
the complex fermion combinations and bosonized representations, $c_{j}=(\psi_{2j-1}+i\psi_{2j})/\sqrt{2}=e^{i\phi^{j}}$
where the bosons obey 
\begin{equation}
\left\langle \phi^{j}\left(z\right)\phi^{j'}\left(w\right)\right\rangle =-\delta^{jj'}\log\left(z-w\right)+\frac{i\pi}{2}\text{sgn}\left(j-j'\right),
\end{equation}
with the sign function accounting for mutual fermionic exchange statistics (Klein factors). We then follow
Reference~\onlinecite{MacfarlaneG2} to embed $G_{2}$ generators
into $SO(7)$. The resulting Cartan generators $H_{G_{2}}^{1,2}$ of $G_{2}$
are
\begin{align}
H_{G_{2}}^{1}\left(z\right)&=i\sqrt{\frac{1}{6}}\left(-2\partial\phi^{1}+\partial\phi^{2}+\partial\phi^{3}\right), \nonumber \\
H_{G_{2}}^{2}\left(z\right)&=i\sqrt{\frac{1}{2}}\left(\partial\phi^{2}-\partial\phi^{3}\right),
\end{align}
while the positive long roots are
\begin{align}
E_{G_{2}}^{1}\left(z\right)&=-e^{i\left(\phi_{2}-\phi_{3}\right)},\nonumber \\
E_{G_{2}}^{2}\left(z\right)&=-e^{i\left(\phi_{3}-\phi_{1}\right)},\nonumber \\
E_{G_{2}}^{3}\left(z\right)&=-e^{i\left(\phi_{2}-\phi_{1}\right)}.
\end{align}

To bosonize the positive short roots, we need to include the fermion $\psi_{7}=\left(e^{i\phi_{4}}+e^{-i\phi_{4}}\right)/\sqrt{2}$,
yielding
\begin{align}\begin{split}
E_{G_{2}}^{4}\left(z\right) & =\frac{1}{\sqrt{3}}\left[-e^{-i\left(\phi_{1}+\phi_{2}\right)}-i\left(e^{i\left(\phi_{3}+\phi_{4}\right)}-e^{i\left(\phi_{3}-\phi_{4}\right)}\right)\right],\\
E_{G_{2}}^{5}\left(z\right) & =\frac{1}{\sqrt{3}}\left[-e^{-i\left(\phi_{1}+\phi_{3}\right)}+i\left(e^{i\left(\phi_{2}+\phi_{4}\right)}-e^{i\left(\phi_{2}-\phi_{4}\right)}\right)\right],\\
E_{G_{2}}^{6}\left(z\right) & =\frac{1}{\sqrt{3}}\left[-e^{i\left(\phi_{2}+\phi_{3}\right)}-i\left(e^{-i\left(\phi_{1}-\phi_{4}\right)}-e^{-i\left(\phi_{1}+\phi_{4}\right)}\right)\right].\end{split}
\end{align}
The negative roots can be obtained by Hermitian conjugation.

Now we move on to $F_{4}$. Our goal is to define the $F_{4}$ currents
in terms of $SO(16)$ degrees of freedom in a way that the operators
decoupled from $G_{2}$, in the operator product expansion (OPE) sense. Since we used the $SO(7)$
part, generated by fermions $\psi_{1,...,7}$ to define the $G_{2}$
operators, we may facilitate the decoupling of the currents by using
the remaining $SO(8)$ subalgebra, generated by $\psi_{9,...,16}$.
This is achieved by carefully sewing $F_{4}$ into the full degrees
of freedom of $SO(16)$. The Cartan generators can be chosen to be
the ones in the $SO(8)$ subalgebra 
\begin{equation}
H_{F_{4}}^{a}(z)=i\partial\phi_{4+a},\,\,a=1,\ldots,4.
\end{equation}

The group $F_{4}$ has 48 roots, 24 short and 24 long. The 24 long
roots are identical to those of $SO(8)$, and may be written in bosonized
form as
\begin{equation}
E_{F_{4}}^{\boldsymbol{\alpha}}(z)=e^{i\boldsymbol{\alpha}\cdot\boldsymbol{\phi}},
\end{equation}
where $\alpha_{1}=\ldots=\alpha_{4}=0$ and $(\alpha_{5},\ldots,\alpha_{8})\in\mathbb{Z}^{4}|\left|(\alpha_{5},\ldots,\alpha_{8})\right|^{2}=2$.
The 24 short roots of $F_{4}$ correspond to 8 vector and 16 spinor
representations of $SO(8)$. To write the 8 vector roots, we increment
the vertex operators with the fermion $\psi_{8}$, obtaining 
\begin{align}
E_{F_{4}}^{\pm a}\sim\psi_{8}e^{\pm i\phi_{4+a}}\sim\frac{1}{\sqrt{2}}\left(e^{i(\phi_{4}\pm\phi_{4+a})}+e^{i(-\phi_{4}\pm\phi_{4+a})}\right).\label{F4shortvector}
\end{align}
Finally, the 16 spinors read 
\begin{align}
E_{F_{4}}^{{\bf s}_{\pm}}\sim\psi_{\pm}e^{i{\bf s}_{\pm}\cdot\boldsymbol{\phi}/2},\label{F4shortspinor}
\end{align}
where the spinor labels are ${\bf s}_{\pm}=(0,0,0,0,s_{5},s_{6},s_{7},s_{8})$
with $s_{5}s_{6}s_{7}s_{8}=\pm1$; the critical step here lies in
the inclusion of the Majorana fermions 
\begin{align}
\psi_{+}&=\frac{1}{\sqrt{2}}\left(\omega_{+}e^{i(\phi_{1}+\phi_{2}+\phi_{3}+\phi_{4})/2}+h.c.\right), \nonumber \\
\psi_{-}&=\frac{1}{\sqrt{2}}\left(\omega_{-}e^{i(\phi_{1}+\phi_{2}+\phi_{3}-\phi_{4})/2}+h.c.\right),\label{MFspinor}
\end{align}
where $\omega_{\pm}$ are $U(1)$ phases to be determined. Combining
the vertices with the fermions, 
\begin{align}
E_{F_{4}}^{{\bf s}_{+}} & \sim\frac{1}{\sqrt{2}}\left(\omega_{+}e^{i(\phi_{1}+\phi_{2}+\phi_{3}+\phi_{4}+{\bf s}_{+}\cdot\boldsymbol{\phi})/2}\right. \nonumber \\
& \quad \quad \quad \quad \quad \quad  \left. +\omega_{+}^{\ast}e^{i(-\phi_{1}-\phi_{2}-\phi_{3}-\phi_{4}+{\bf s}_{+}\cdot\boldsymbol{\phi})/2}\right), \nonumber \\
E_{F_{4}}^{{\bf s}_{-}} & \sim\frac{i}{\sqrt{2}}\left(\omega_{-}e^{i(\phi_{1}+\phi_{2}+\phi_{3}-\phi_{4}+{\bf s}_{-}\cdot\boldsymbol{\phi})/2} \right. \nonumber \\
 & \quad \quad \quad \quad \quad \quad  \left. - \omega_{-}^{\ast}e^{i(-\phi_{1}-\phi_{2}-\phi_{3}+\phi_{4}+{\bf s}_{-}\cdot\boldsymbol{\phi})/2}\right).
 \end{align}

Our goal is to decouple the $G_{2}$ and $F_{4}$ currents in the
SO(16) embedding. Computing the OPEs between all $G_{2}$ and $F_{4}$
operators, one recognizes that singular terms only arise between $G_{2}$
short roots and $F_{4}$ short roots from $SO(8)$ spinors. These
singular terms, however, can be made to vanish with an appropriate
choice of $\omega_{\pm}$ following 
\begin{equation}
\omega_{+}+e^{-i\pi/4}\omega_{+}^{*}=\omega_{-}-e^{-i\pi/4}\omega_{-}^{*}=0.
\end{equation}
Distinct solutions only differ by a sign, which can be absorbed in
the Majorana fermion $\psi_{\pm}$. We pick 
\begin{align}
\omega_{+}=e^{i3\pi/8},\quad\omega_{-}=e^{-i\pi/8}.\label{omega}
\end{align}
This completes the proof that the $G_{2}$ and $F_{4}$ embeddings
decouple and act on distinct Hilbert spaces.

Besides the OPE decomposition, as a non-trivial complementary check of the conformal embedding involves the computation of energy-momentum tensors, seeing that the $E_{8}$ tensor decouples
identically into those of $G_{2}$ and $F_{4}$ under the construction
above. The calculation is possible, albeit involved; the results are presented in Appendix~\ref{subsec:G2F4_EM_tensor}.

\begin{figure}[t!]
\centering\includegraphics[width=0.8\columnwidth]{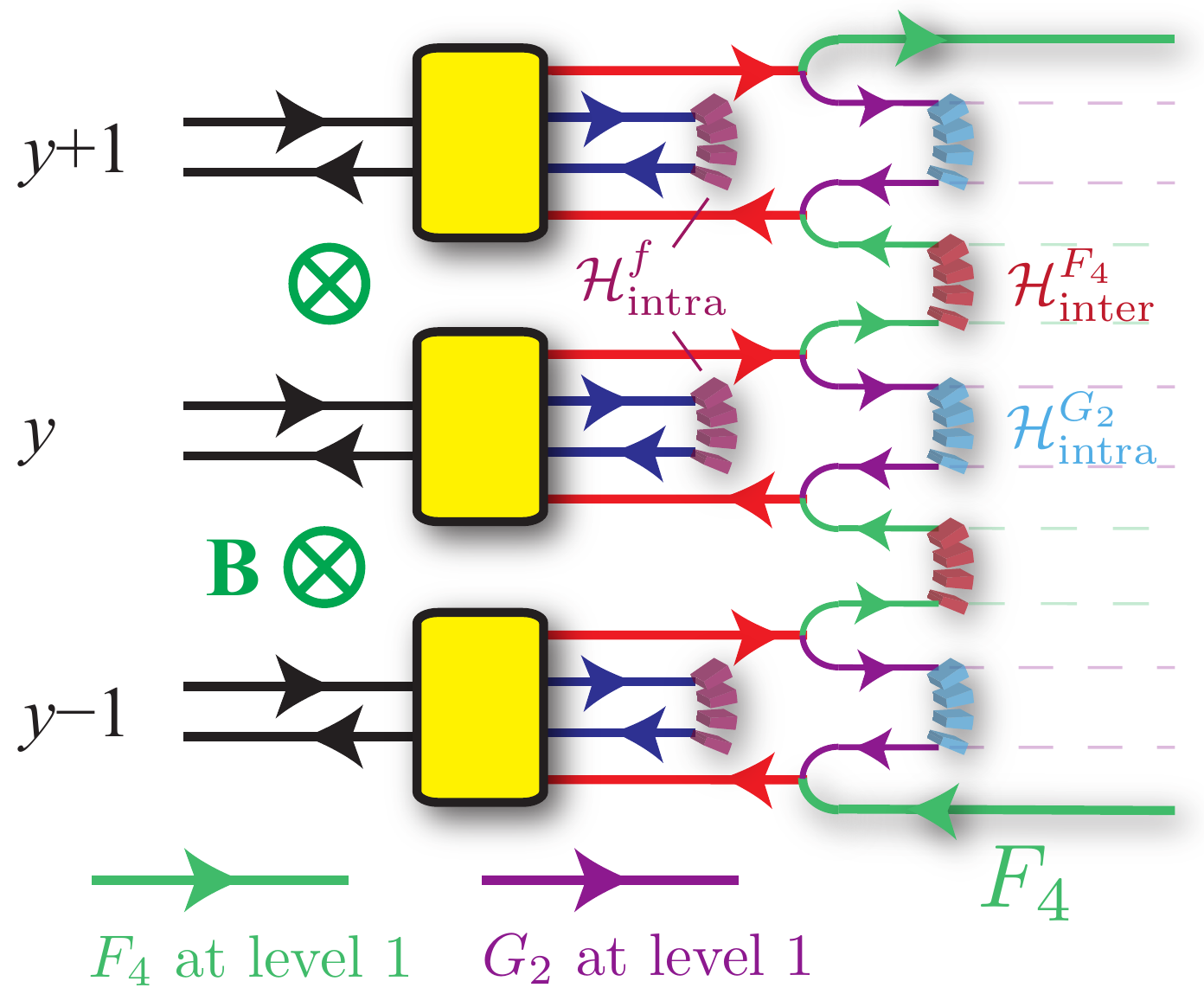}
\caption{The coupled wire model \eqref{eq:F4back} for the $F_{4}$ Fibonacci
quantum Hall state at filling $\nu=8$.}
\label{fig:G2F4couplewire} 
\end{figure}

\subsection{ \texorpdfstring{$G_{2}$}{G2} and \texorpdfstring{$ F_{4}$}{F4} Fibonacci topological order via coupled-wires }

We have all the structure necessary for the coupled-wire construction of the $F_4$ and $G_2$ phases.
Similar to the $E_{8}$ state, these quantum Hall phases are based on an array of 11-wire
bundles. Fig.~\ref{fig:G2F4couplewire} shows the schematics of the backscattering
terms in the $F_{4}$ quantum Hall Hamiltonian case. The $G_{2}$ state
can be described using a similar diagram by switching the roles of
$G_{2}$ and $F_{4}$. The models are written with the intra-bundle backscattering \eqref{eq:IntraU1},
which leaves behind a counter-propagating pair of $E_{8}$ modes per
bundle. The $\mathcal{G}=F_{4}$ or $G_{2}$ currents are then dimerized
within or between bundles according to 
\begin{align}
\begin{split}\mathcal{H}_{\mathrm{intra}}^{y,\mathcal{G}} & =u_{\mathrm{intra}}\sum_{\boldsymbol{\alpha}\in\Delta_{\mathcal{G}}}{\left[E_{\mathcal{G}}\right]_{y,\boldsymbol{\alpha}}^{R}}^{\dagger}\left[E_{\mathcal{G}}\right]_{y,\boldsymbol{\alpha}}^{L}+h.c.,\\
\mathcal{H}_{\mathrm{inter}}^{y+1/2,\mathcal{G}} & =u_{\mathrm{inter}}\sum_{\boldsymbol{\alpha}\in\Delta_{\mathcal{G}}}{\left[E_{\mathcal{G}}\right]_{y,\boldsymbol{\alpha}}^{R}}^{\dagger}\left[E_{\mathcal{G}}\right]_{y+1,\boldsymbol{\alpha}}^{L}+h.c.~.
\end{split}
\label{Gback}
\end{align}
The $F_{4}$ and $G_{2}$ quantum Hall states consist, respectively,
of the ground states of the following Hamiltonians,
\begin{align}
\mathcal{H}[F_{4}] & =\sum_{y=1}^{N_{l}}\left(\mathcal{H}_{\mathrm{intra}}^{y,f}+\mathcal{H}_{\mathrm{intra}}^{y,G_{2}}\right)+\sum_{y=1}^{N_{l}-1}\mathcal{H}_{\mathrm{inter}}^{y+1/2,F_{4}},\label{eq:F4back}\\
\mathcal{H}[G_{2}] & =\sum_{y=1}^{N_{l}}\left(\mathcal{H}_{\mathrm{intra}}^{y,f}+\mathcal{H}_{\mathrm{intra}}^{y,F_{4}}\right)+\sum_{y=1}^{N_{l}-1}\mathcal{H}_{\mathrm{inter}}^{y+1/2,G_{2}}.\label{eq:G2back}
\end{align}

The momentum-conservation conditions have to be reimplemented to the many-body interactions in either \eqref{eq:F4back} or \eqref{eq:G2back}. Each phase is stabilized by its own distribution of electronic momenta $k_{ya}^{\sigma}$~(c.f. Appendix~\ref{subsec:Fermi-momenta-and-filling-G2-F4}), but both have the \emph{same} integer magnetic filling $\nu=8$. At strong coupling, $\mathcal{H}[F_{4}]$ ($\mathcal{H}[G_{2}]$) gives rise to a finite excitation energy gap in the bulk, but leaves behind a gapless chiral $F_{4}$ ($G_{2}$) \hyperlink{WZW}{WZW} \hyperlink{CFT}{CFT} at level 1 at the boundary. As a consequence, the Wiedemann-Franz law is again unconventional in these phases, displaying $c_{F_4}/\nu=13/20$ and $c_{G_2}/\nu=7/20$.

According to the bulk-boundary correspondence, the anyon content of the $F_4$ and $G_2$ phases can be read from their boundary theories. In Appendix~\ref{subsec:Fibonacci-primary-fields} we present an extensive discussion about the relationship of these phases with Fibonacci topological order; here we just describe the general facts. In addition to the vacuum $1$, each edge carries a Fibonacci primary field $\bar{\tau}$ for $(F_{4})_{1}$ and $\tau$ for $(G_{2})_{1}$, with conformal scaling dimensions $3/5$ and $2/5$ respectively. Each consists of a collection of operators, known as a super-selection sector, that corresponds to the 26 dimensional (7 dimensional) fundamental representation of $F_{4}$ ($G_{2}$) that rotates under the \hyperlink{WZW}{WZW} algebra. Our construction allows an explicit parafermionic representation of these fields (see Appendix~\ref{subsec:Fibonacci-primary-fields}). Here, we notice that since the current operators $[E_{F_{4}}]_{\boldsymbol{\alpha}}$ are even combinations of electrons, the Fibonacci operators within a super-sector differ from each other by pairs of electrons, and therefore correspond to the same anyon type. Moreover, they all have even electric charge and therefore the gapless chiral edge \hyperlink{CFT}{CFT} only supports even charge low-energy excitations. An analogous analysis follows for the $G_2$ case.


\subsection{ Particle-hole conjugation and Fibonacci vs anti-Fibonacci phases }

As our final comments, notice that the $G_{2}$ and $F_{4}$ Fibonacci states at $\nu=8$ half-fill the $E_{8}$ quantum Hall state, which has $\nu=16$. Remarkably, they are related under a notion of particle-hole ({\color{blue}\hypertarget{PH}{PH}}) conjugation that is based on $E_{8}$ bosons instead of electrons. A similar generalization of \hyperlink{PH}{PH} symmetry has been proposed for parton quantum Hall states~\cite{SirotaSahooChoTeo18}.
The \hyperlink{PH}{PH} conjugation manifests in the edge \hyperlink{CFT}{CFT} as the coset identities $(G_{2})_{1}=(E_{8})_{1}/(F_{4})_{1}$ and $(F_{4})_{1}=(E_{8})_{1}/(G_{2})_{1}$, which reflect the equality $T_{G_{2}}+T_{F_{4}}=T_{E_{8}}$ between energy-momentum tensors. The coset $E_{8}/\mathcal{G}$
can be understood as the subtraction of the \hyperlink{WZW}{WZW} sub-algebra $\mathcal{G}$ from $E_{8}$. In other words, this is equivalent to the tensor product $E_{8}\otimes\overline{\mathcal{G}}$, where the time-reversal conjugate $\overline{\mathcal{G}}$ pair annihilates with the \hyperlink{WZW}{WZW} sub-algebra $\mathcal{G}$ in $E_{8}$ by current-current backscattering interactions similar to \eqref{Gback}.  The coset identities are direct consequences of the conformal embedding $(G_{2})_{1}\times(F_{4})_{1}\subseteq(E_{8})_{1}$.

The conventional \hyperlink{PH}{PH} symmetry of the half-filled Landau level has been studied in the coupled wire context~\cite{MrossAliceaMotrunich16,MrossAliceaMotrunich16PRL,MrossAliceaMotrunich17,FujiFurusaki18}. Here, the $E_{8}$-based \hyperlink{PH}{PH} conjugation has a microscopic description as well. It is represented by an anti-unitary operator $\mathcal{C}$ that relates the $E_{8}$ bosonized variables between the two Fibonacci states 
\begin{align}
\begin{split}\mathcal{C}\tilde{\Phi}_{y,I}^{R}\mathcal{C}^{-1} & =\tilde{\Phi}_{y,I}^{L}-q_{I}\mathsf{x}/2\\
\mathcal{C}\tilde{\Phi}_{y,I}^{L}\mathcal{C}^{-1} & =\tilde{\Phi}_{y-1,I}^{R}-q_{I}\mathsf{x}/2
\end{split}
\label{PHconjugation}
\end{align}
while leaving the recombined Dirac fermions unaltered, $\mathcal{C}f_{yn}^{\sigma}\mathcal{C}^{-1}=f_{yn}^{\sigma}$. Since the $E_{8}$ root structure is unimodular the \hyperlink{PH}{PH} conjugation \eqref{PHconjugation} is an integral action of the fundamental electrons, $\mathcal{C}c_{J}\mathcal{C}^{-1}=\prod_{J'}(c_{J'})^{m_{J'}^{J}}$, where $m_{J'}^{J}$ are integers, $J,J'$ are the collections of indices $y,a,\sigma$, and the product is finite and short-ranged so that it only involves nearest neighboring bundles $|y-y'|\leq1$. The \hyperlink{PH}{PH} conjugation switches between intra- and inter-bundle interactions of the $G_{2}$ and $F_{4}$ currents, exchanging the two Fibonacci phases $\mathcal{C}\mathcal{H}[F_{4}]\mathcal{C}^{-1}=\mathcal{H}[G_{2}]$ and $\mathcal{C}\mathcal{H}[G_{2}]\mathcal{C}^{-1}=\mathcal{H}[F_{4}]$. Lastly, the coupled wire description artificially causes the \hyperlink{PH}{PH} conjugation to be non-local. Similar to an antiferromagnetic symmerty, $\mathcal{C}^{2}$ unitarily translates the $E_{8}$ currents from $y$ to $y-1$.

\section{Conclusions}\label{sec:conclusions}

We presented a coupled-wire construction based on exceptional Lie algebras of three distinct time-reversal broken topological phases carrying bosonic edge modes. The first one, the $E_8$ quantum Hall state, displays short-ranged entanglement. The other two phases are long-range entangled, non-Abelian, and based on the $G_2$ and $F_4$ algebras. These latter two define Fibonacci topological ordered states. Crucially, all of these phases are predicted to exist within integer magnetic filling fractions, $\nu=16$ for $E_8$ and $\nu=8$ for the $G_2$ and $F_4$, suggesting that interactions inside integer quantum Hall plateaus may be used to stabilize these phases. 

Our findings are allowed by technical advances we introduce regarding the representation of more complex current algebras in the coupled-wire program. This microscopic approach proves to go beyond the standard Chern-Simons effective field theory of topological phases,  allowing us to  settle down a concrete system where the $E_8$ state which may be pursued, namely the $\nu=16$ integer quantum Hall plateau. The method also allows the extra prediction of non-Abelian Fibonacci topological ordered phases in integer Hall plateaus, as well as the definition of a particle-hole operation that connects the Fibonacci and anti-Fibonacci phases. These results are of practical relevance, given the importance of Fibonacci anyons in topological quantum computation by anyons.

The most evident phenomenological distinction of the $E_{8}$, $G_{2}$ and $F_{4}$ states, as we argued, stems from modified Wiedemann-Franz laws, with distinct $c/\nu$ ratios. The presence  of these phases in low temperature at filling $\nu=16$ or 8 could be verified by thermal Hall transport measurements. Similar thermal
conductance observations have recently been recently performed for other fractional quantum Hall states~\cite{BanerjeeHeiblumRosenblattOregFeldmanSternUmansky17,BanerjeeHeiblumUmanskyFeldmanOregStern18}. Moreover, all three quantum Hall states here proposed carry bosonic edge modes that only support even charge gapless quasiparticles. This gives rise to a distinct shot noise signature across a point contact below the energy gap. The anyonic statistics of the Fibonacci excitations in the $G_{2}$ and $F_{4}$ states can be detected by Fabry-Perot interferometry. 

Another relevant question, which will be saved for future inquiries, regards pinpointing specific interactions leading to these phases at an actual quantum Hall electron fluid setting. We believe a variational wavefunctional approach, similar to Laughlin's construction of his wavefunctions for fractional quantum Hall systems might be a promising approach. Finally, due to thermal fluctuations and disorder, Hall plateaus as high as $\nu=8$ or $16$ are challenging, albeit not impossible, to probe experimentally. While nothing precludes such measurements this fundamentally, a promising future path of inquiry lies in also searching for other topological phase transitions in lower, and more stable, magnetic fillings. A guiding principle for this search involves searching phases where $c$ and $\nu$ are different mod 8~\cite{Kitaev06}. 


\begin{acknowledgments}
We thank Yichen Hu and Charlie Kane for insightful discussion. JCYT
is supported by the National Science Foundation under Grant No.~DMR-1653535.
PLSL is supported by the Canada First Research Excellence Fund and NSERC. VLQ acknowledges support
by the NSF Grant No. DMR-1644779 and the Aspen Center for Physics, under 
the NSF grant PHY-1607611, for hospitality. BH acknowledges support from the ICMT at UIUC.
\end{acknowledgments}


\begin{thebibliography}{41}
\makeatletter
\providecommand \@ifxundefined [1]{%
 \@ifx{#1\undefined}
}%
\providecommand \@ifnum [1]{%
 \ifnum #1\expandafter \@firstoftwo
 \else \expandafter \@secondoftwo
 \fi
}%
\providecommand \@ifx [1]{%
 \ifx #1\expandafter \@firstoftwo
 \else \expandafter \@secondoftwo
 \fi
}%
\providecommand \natexlab [1]{#1}%
\providecommand \enquote  [1]{``#1''}%
\providecommand \bibnamefont  [1]{#1}%
\providecommand \bibfnamefont [1]{#1}%
\providecommand \citenamefont [1]{#1}%
\providecommand \href@noop [0]{\@secondoftwo}%
\providecommand \href [0]{\begingroup \@sanitize@url \@href}%
\providecommand \@href[1]{\@@startlink{#1}\@@href}%
\providecommand \@@href[1]{\endgroup#1\@@endlink}%
\providecommand \@sanitize@url [0]{\catcode `\\12\catcode `\$12\catcode
  `\&12\catcode `\#12\catcode `\^12\catcode `\_12\catcode `\%12\relax}%
\providecommand \@@startlink[1]{}%
\providecommand \@@endlink[0]{}%
\providecommand \url  [0]{\begingroup\@sanitize@url \@url }%
\providecommand \@url [1]{\endgroup\@href {#1}{\urlprefix }}%
\providecommand \urlprefix  [0]{URL }%
\providecommand \Eprint [0]{\href }%
\providecommand \doibase [0]{http://dx.doi.org/}%
\providecommand \selectlanguage [0]{\@gobble}%
\providecommand \bibinfo  [0]{\@secondoftwo}%
\providecommand \bibfield  [0]{\@secondoftwo}%
\providecommand \translation [1]{[#1]}%
\providecommand \BibitemOpen [0]{}%
\providecommand \bibitemStop [0]{}%
\providecommand \bibitemNoStop [0]{.\EOS\space}%
\providecommand \EOS [0]{\spacefactor3000\relax}%
\providecommand \BibitemShut  [1]{\csname bibitem#1\endcsname}%
\let\auto@bib@innerbib\@empty
\bibitem [{\citenamefont {Wilczek}(1990)}]{Wilczekbook}%
  \BibitemOpen
  \bibfield  {author} {\bibinfo {author} {\bibfnamefont {F.}~\bibnamefont
  {Wilczek}},\ }\href@noop {} {\emph {\bibinfo {title} {Fractional Statistics
  and Anyon Superconductivity}}}\ (\bibinfo  {publisher} {World Scientific},\
  \bibinfo {year} {1990})\BibitemShut {NoStop}%
\bibitem [{\citenamefont {Wen}(2004)}]{Wenbook}%
  \BibitemOpen
  \bibfield  {author} {\bibinfo {author} {\bibfnamefont {X.-G.}\ \bibnamefont
  {Wen}},\ }\href@noop {} {\emph {\bibinfo {title} {Quantum Field Theory of
  Many Body Systems}}}\ (\bibinfo  {publisher} {Oxford Univ. Press, Oxford},\
  \bibinfo {year} {2004})\BibitemShut {NoStop}%
\bibitem [{\citenamefont {Nayak}\ \emph {et~al.}(2008)\citenamefont {Nayak},
  \citenamefont {Simon}, \citenamefont {Stern}, \citenamefont {Freedman},\ and\
  \citenamefont {Das~Sarma}}]{ChetanSimonSternFreedmanDasSarma}%
  \BibitemOpen
  \bibfield  {author} {\bibinfo {author} {\bibfnamefont {C.}~\bibnamefont
  {Nayak}}, \bibinfo {author} {\bibfnamefont {S.~H.}\ \bibnamefont {Simon}},
  \bibinfo {author} {\bibfnamefont {A.}~\bibnamefont {Stern}}, \bibinfo
  {author} {\bibfnamefont {M.}~\bibnamefont {Freedman}}, \ and\ \bibinfo
  {author} {\bibfnamefont {S.}~\bibnamefont {Das~Sarma}},\ }\href {\doibase
  10.1103/RevModPhys.80.1083} {\bibfield  {journal} {\bibinfo  {journal} {Rev.
  Mod. Phys.}\ }\textbf {\bibinfo {volume} {80}},\ \bibinfo {pages} {1083}
  (\bibinfo {year} {2008})}\BibitemShut {NoStop}%
\bibitem [{\citenamefont {Cage}\ \emph {et~al.}(2012)\citenamefont {Cage},
  \citenamefont {Klitzing}, \citenamefont {Chang}, \citenamefont {Duncan},
  \citenamefont {Haldane}, \citenamefont {Laughlin}, \citenamefont {Pruisken},
  \citenamefont {Thouless}, \citenamefont {Prange},\ and\ \citenamefont
  {Girvin}}]{FQHE_Review}%
  \BibitemOpen
  \bibfield  {author} {\bibinfo {author} {\bibfnamefont {M.~E.}\ \bibnamefont
  {Cage}}, \bibinfo {author} {\bibfnamefont {K.}~\bibnamefont {Klitzing}},
  \bibinfo {author} {\bibfnamefont {A.}~\bibnamefont {Chang}}, \bibinfo
  {author} {\bibfnamefont {F.}~\bibnamefont {Duncan}}, \bibinfo {author}
  {\bibfnamefont {M.}~\bibnamefont {Haldane}}, \bibinfo {author} {\bibfnamefont
  {R.}~\bibnamefont {Laughlin}}, \bibinfo {author} {\bibfnamefont
  {A.}~\bibnamefont {Pruisken}}, \bibinfo {author} {\bibfnamefont
  {D.}~\bibnamefont {Thouless}}, \bibinfo {author} {\bibfnamefont {R.~E.}\
  \bibnamefont {Prange}}, \ and\ \bibinfo {author} {\bibfnamefont {S.~M.}\
  \bibnamefont {Girvin}},\ }\href@noop {} {\emph {\bibinfo {title} {The Quantum
  Hall Effect}}}\ (\bibinfo  {publisher} {Springer Science \& Business Media,
  Berlin},\ \bibinfo {year} {2012})\BibitemShut {NoStop}%
\bibitem [{\citenamefont {Halperin}\ \emph {et~al.}(1993)\citenamefont {Halperin},
  \citenamefont {Lee},\ and\ \citenamefont
  {Read}}]{HLR_composite}%
  \BibitemOpen
  \bibfield  {author} {\bibinfo {author} {\bibfnamefont {B.~I.}\ \bibnamefont
  {Halperin}}, \bibinfo {author} {\bibfnamefont {P.~A.}~\bibnamefont {Lee}},
  \ and\ \bibinfo {author} {\bibfnamefont {N.}\ \bibnamefont {Read}},\
  }\href {\doibase 10.1103/PhysRevB.47.7312} {\bibfield  {journal}
  {\bibinfo  {journal} {Phys. Rev. B}\ }\textbf {\bibinfo {volume} {47}},\
  \bibinfo {pages} {7312} (\bibinfo {year} {1993})}\BibitemShut {NoStop}%
\bibitem [{\citenamefont {Son}\ \emph {D.~T.}(2015)\citenamefont {Son},
  \citenamefont {Dam}}]{Son_composite}%
  \BibitemOpen
  \bibfield  {author} {\bibinfo {author} {\bibfnamefont {B.~I.}\ \bibnamefont
  {Halperin}}, \bibinfo {author} {\bibfnamefont {P.~A.}~\bibnamefont {Lee}},
  \ and\ \bibinfo {author} {\bibfnamefont {N.}\ \bibnamefont {Read}},\
  }\href {\doibase 10.1103/PhysRevB.47.7312} {\bibfield  {journal}
  {\bibinfo  {journal} {Phys. Rev. B}\ }\textbf {\bibinfo {volume} {47}},\
  \bibinfo {pages} {7312} (\bibinfo {year} {1993})}\BibitemShut {NoStop}%
\bibitem [{\citenamefont {Moore}\ and\ \citenamefont {Read}(1991)}]{MooreRead}%
  \BibitemOpen
  \bibfield  {author} {\bibinfo {author} {\bibfnamefont {G.}~\bibnamefont
  {Moore}}\ and\ \bibinfo {author} {\bibfnamefont {N.}\ \bibnamefont {Read}},\
  }\href {\doibase /10.1016/0550-3213(91)90407-O} {\bibfield  {journal}
  {\bibinfo  {journal} {Nucl. Phys. B}\ }\textbf {\bibinfo {volume} {360}},\
  \bibinfo {pages} {362} (\bibinfo {year} {1991})}\BibitemShut {NoStop}%
\bibitem [{\citenamefont {Lu}\ and\ \citenamefont
  {Vishwanath}(2012)}]{LuVish}%
  \BibitemOpen
  \bibfield  {author} {\bibinfo {author} {\bibfnamefont {Y.-M.}\ \bibnamefont
  {Lu}}\ and\ \bibinfo {author} {\bibfnamefont {A.}\ \bibnamefont
  {Vishwanath}},\ }\href {\doibase 10.1103/PhysRevB.86.125119} {\bibfield
  {journal} {\bibinfo  {journal} {Phys. Rev. B}\ }\textbf {\bibinfo
  {volume} {86}},\ \bibinfo {pages} {125119} (\bibinfo {year}
  {2012})}\BibitemShut {NoStop}%
\bibitem [{\citenamefont {Plamadeala}\ \emph {et~al.}(2013)\citenamefont {Plamadeala},
  \citenamefont {Mulligan},\ and\ \citenamefont
  {Nayak}}]{Plama}%
  \BibitemOpen
  \bibfield  {author} {\bibinfo {author} {\bibfnamefont {E.}\ \bibnamefont
  {Plamadeala}}, \bibinfo {author} {\bibfnamefont {M.}~\bibnamefont {Mulligan}},
  \ and\ \bibinfo {author} {\bibfnamefont {C.}\ \bibnamefont {Nayak}},\
  }\href {\doibase 10.1103/PhysRevB.88.045131} {\bibfield  {journal}
  {\bibinfo  {journal} {Phys. Rev. B}\ }\textbf {\bibinfo {volume} {88}},\
  \bibinfo {pages} {045131} (\bibinfo {year} {2013})}\BibitemShut {NoStop}%
\bibitem [{\citenamefont {Kane}\ \emph {et~al.}(2002)\citenamefont {Kane},
  \citenamefont {Mukhopadhyay},\ and\ \citenamefont
  {Lubensky}}]{KaneMukhopadhyayLubensky02}%
  \BibitemOpen
  \bibfield  {author} {\bibinfo {author} {\bibfnamefont {C.~L.}\ \bibnamefont
  {Kane}}, \bibinfo {author} {\bibfnamefont {R.}~\bibnamefont {Mukhopadhyay}},
  \ and\ \bibinfo {author} {\bibfnamefont {T.~C.}\ \bibnamefont {Lubensky}},\
  }\href {\doibase 10.1103/PhysRevLett.88.036401} {\bibfield  {journal}
  {\bibinfo  {journal} {Phys. Rev. Lett.}\ }\textbf {\bibinfo {volume} {88}},\
  \bibinfo {pages} {036401} (\bibinfo {year} {2002})}\BibitemShut {NoStop}%
\bibitem [{\citenamefont {Sahoo}\ \emph {et~al.}(2013)\citenamefont {Sahoo},
  \citenamefont {Zhang},\ and\ \citenamefont
  {Teo}}]{Sahoo16}%
  \BibitemOpen
  \bibfield  {author} {\bibinfo {author} {\bibfnamefont {S.}\ \bibnamefont
  {Sahoo}}, \bibinfo {author} {\bibfnamefont {Z.}~\bibnamefont {Zhang}},
  \ and\ \bibinfo {author} {\bibfnamefont {J. C. Y.}\ \bibnamefont {Teo}},\
  }\href {\doibase 10.1103/PhysRevB.94.165142} {\bibfield  {journal}
  {\bibinfo  {journal} {Phys. Rev. B}\ }\textbf {\bibinfo {volume} {94}},\
  \bibinfo {pages} {165142} (\bibinfo {year} {2016})}\BibitemShut {NoStop}%
\bibitem [{\citenamefont {Teo}\ and\ \citenamefont
  {Kane}(2014)}]{TeoKaneCW}%
  \BibitemOpen
  \bibfield  {author} {\bibinfo {author} {\bibfnamefont {J. C. Y.}\ \bibnamefont
  {Teo}}\ and\ \bibinfo {author} {\bibfnamefont {C. L.}\ \bibnamefont
  {Kane}},\ }\href {\doibase 10.1103/PhysRevB.89.085101} {\bibfield
  {journal} {\bibinfo  {journal} {Phys. Rev. B}\ }\textbf {\bibinfo
  {volume} {89}},\ \bibinfo {pages} {085101} (\bibinfo {year}
  {2014})}\BibitemShut {NoStop}%
\bibitem [{\citenamefont {Franz}\ and\ \citenamefont
  {Wiedemann}(1853)}]{FranzWiedemann}%
  \BibitemOpen
  \bibfield  {author} {\bibinfo {author} {\bibfnamefont {R.}~\bibnamefont
  {Franz}}\ and\ \bibinfo {author} {\bibfnamefont {G.}~\bibnamefont
  {Wiedemann}},\ }\href {\doibase 10.1002/andp.18531650802} {\bibfield
  {journal} {\bibinfo  {journal} {Annalen der Physik}\ }\textbf {\bibinfo
  {volume} {165}},\ \bibinfo {pages} {497} (\bibinfo {year}
  {1853})}\BibitemShut {NoStop}%
\bibitem [{\citenamefont {Trebst}\ \emph {et~al.}(2008)\citenamefont {Trebst},
  \citenamefont {Troyer}, \citenamefont {Wang},\ and\ \citenamefont
  {Ludwig}}]{TrebstTroyerWangLudwig08}%
  \BibitemOpen
  \bibfield  {author} {\bibinfo {author} {\bibfnamefont {S.}~\bibnamefont
  {Trebst}}, \bibinfo {author} {\bibfnamefont {M.}~\bibnamefont {Troyer}},
  \bibinfo {author} {\bibfnamefont {Z.}~\bibnamefont {Wang}}, \ and\ \bibinfo
  {author} {\bibfnamefont {A.~W.~W.}\ \bibnamefont {Ludwig}},\ }\href {\doibase
  10.1143/PTPS.176.384} {\bibfield  {journal} {\bibinfo  {journal} {Progress of
  Theoretical Physics Supplement}\ }\textbf {\bibinfo {volume} {176}},\
  \bibinfo {pages} {384} (\bibinfo {year} {2008})}\BibitemShut {NoStop}%
\bibitem [{\citenamefont {Read}\ and\ \citenamefont
  {Rezayi}(1999)}]{ReadRezayi}%
  \BibitemOpen
  \bibfield  {author} {\bibinfo {author} {\bibfnamefont {N.}~\bibnamefont
  {Read}}\ and\ \bibinfo {author} {\bibfnamefont {E.}~\bibnamefont {Rezayi}},\
  }\href {\doibase 10.1103/PhysRevB.59.8084} {\bibfield  {journal} {\bibinfo
  {journal} {Phys. Rev. B.}\ }\textbf {\bibinfo {volume} {59}},\ \bibinfo
  {pages} {8084} (\bibinfo {year} {1999})}\BibitemShut {NoStop}%
\bibitem [{\citenamefont {Mong}\ \emph {et~al.}(2014)\citenamefont {Mong},
  \citenamefont {Clarke}, \citenamefont {Alicea}, \citenamefont {Lindner},
  \citenamefont {Fendley}, \citenamefont {Nayak}, \citenamefont {Oreg},
  \citenamefont {Stern}, \citenamefont {Berg}, \citenamefont {Shtengel},\ and\
  \citenamefont {Fisher}}]{mongg2}%
  \BibitemOpen
  \bibfield  {author} {\bibinfo {author} {\bibfnamefont {R.~S.}\ \bibnamefont
  {Mong}}, \bibinfo {author} {\bibfnamefont {D.~J.}\ \bibnamefont {Clarke}},
  \bibinfo {author} {\bibfnamefont {J.}~\bibnamefont {Alicea}}, \bibinfo
  {author} {\bibfnamefont {N.~H.}\ \bibnamefont {Lindner}}, \bibinfo {author}
  {\bibfnamefont {P.}~\bibnamefont {Fendley}}, \bibinfo {author} {\bibfnamefont
  {C.}~\bibnamefont {Nayak}}, \bibinfo {author} {\bibfnamefont
  {Y.}~\bibnamefont {Oreg}}, \bibinfo {author} {\bibfnamefont {A.}~\bibnamefont
  {Stern}}, \bibinfo {author} {\bibfnamefont {E.}~\bibnamefont {Berg}},
  \bibinfo {author} {\bibfnamefont {K.}~\bibnamefont {Shtengel}}, \ and\
  \bibinfo {author} {\bibfnamefont {M.~P.}\ \bibnamefont {Fisher}},\ }\href
  {\doibase 10.1103/PhysRevX.4.011036} {\bibfield  {journal} {\bibinfo
  {journal} {Phys. Rev. X}\ }\textbf {\bibinfo {volume} {4}},\ \bibinfo {pages}
  {011036} (\bibinfo {year} {2014})}\BibitemShut {NoStop}%
\bibitem [{\citenamefont {Hu}\ and\ \citenamefont {Kane}(2018)}]{HuKane18}%
  \BibitemOpen
  \bibfield  {author} {\bibinfo {author} {\bibfnamefont {Y.}~\bibnamefont
  {Hu}}\ and\ \bibinfo {author} {\bibfnamefont {C.~L.}\ \bibnamefont {Kane}},\
  }\href {\doibase 10.1103/PhysRevLett.120.066801} {\bibfield  {journal}
  {\bibinfo  {journal} {Phys. Rev. Lett.}\ }\textbf {\bibinfo {volume} {120}},\
  \bibinfo {pages} {066801} (\bibinfo {year} {2018})}\BibitemShut {NoStop}%
\bibitem [{\citenamefont {Kitaev}(2006)}]{Kitaev06}%
  \BibitemOpen
  \bibfield  {author} {\bibinfo {author} {\bibfnamefont {A.}~\bibnamefont
  {Kitaev}}} {\bibfield  {journal}
  {\bibinfo  {journal} {Ann. Phys.}\ }\textbf {\bibinfo {volume} {321}},\
  (\bibinfo {year} {2014})}\BibitemShut {NoStop}%
\bibitem [{\citenamefont {Wen}(2006)}]{Wen1990}%
  \BibitemOpen
  \bibfield  {author} {\bibinfo {author} {\bibfnamefont {X.-G.}~\bibnamefont
  {Wen}}} {\bibfield  {journal}
  {\bibinfo  {journal} {Int. J. Mod. Phys. B}\ }\textbf {\bibinfo {volume} {4}},\
  (\bibinfo {year} {1990})}\BibitemShut {NoStop}%
\bibitem [{\citenamefont {Furey}(2014)}]{Furey14}%
  \BibitemOpen
  \bibfield  {author} {\bibinfo {author} {\bibfnamefont {C.}~\bibnamefont
  {Furey}}} {\bibfield  {journal}
  {\bibinfo  {journal} {JHEP}\ }\textbf {\bibinfo {volume} {46}},\
  \bibinfo {issue} {11} (\bibinfo {year} {2014})}\BibitemShut {NoStop}%
\bibitem [{\citenamefont {Zamolodchikov}(1989)}]{ZamolodchikovE8}%
  \BibitemOpen
  \bibfield  {author} {\bibinfo {author} {\bibfnamefont {A.}~\bibnamefont
  {Zamolodchikov}}} {\bibfield  {journal}
  {\bibinfo  {journal} {Int. J. of Mod. Phys. A}\ }\textbf {\bibinfo {volume} {4}},\
  \bibinfo {issue} {16} (\bibinfo {year} {1989})}\BibitemShut {NoStop}%
\bibitem [{\citenamefont {Coldea}\ \emph {et~al.}(2010)\citenamefont {Coldea},
  \citenamefont {Tennant}, \citenamefont {Wheeler}, \citenamefont {Wawrzynska},
  \citenamefont {Prabhakaran}, \citenamefont {Telling}, \citenamefont {Habicht},
  \citenamefont {Smeibidil}, \citenamefont {Kiefer}}]{E8Science}%
  \BibitemOpen
  \bibfield  {author} {\bibinfo {author} {\bibfnamefont {R.}\ \bibnamefont
  {Coldea}}, \bibinfo {author} {\bibfnamefont {D.~A.}\ \bibnamefont {Tennant}},
  \bibinfo {author} {\bibfnamefont {E.~M.}~\bibnamefont {Wheeler}}, \bibinfo
  {author} {\bibfnamefont {E.}\ \bibnamefont {Wawrzynska}}, \bibinfo {author}
  {\bibfnamefont {D.}~\bibnamefont {Prabhakaran}}, \bibinfo {author} {\bibfnamefont
  {M.}~\bibnamefont {Telling}}, \bibinfo {author} {\bibfnamefont
  {K.}~\bibnamefont {Habicht}}, \bibinfo {author} {\bibfnamefont {P.}~\bibnamefont
  {Smeibidil}}, \bibinfo {author} {\bibfnamefont {K.}~\bibnamefont {Kiefer}},\ }\href
  {\doibase 10.1126/science.1180085} {\bibfield  {journal} {\bibinfo
  {journal} {Science}\ }\textbf {\bibinfo {volume} {8}},\ \bibinfo {issue}
  {5962} (\bibinfo {year} {2010})}\BibitemShut {NoStop}%
\bibitem [{\citenamefont {Rowell}\ \emph {et~al.}(2009)\citenamefont {Rowell},
  \citenamefont {Stong},\ and\ \citenamefont {Wang}}]{RowellStongWang09}%
  \BibitemOpen
  \bibfield  {author} {\bibinfo {author} {\bibfnamefont {E.}~\bibnamefont
  {Rowell}}, \bibinfo {author} {\bibfnamefont {R.}~\bibnamefont {Stong}}, \
  and\ \bibinfo {author} {\bibfnamefont {Z.}~\bibnamefont {Wang}},\ }\href
  {\doibase 10.1007/s00220-009-0908-z} {\bibfield  {journal} {\bibinfo
  {journal} {Communications in Mathematical Physics}\ }\textbf {\bibinfo
  {volume} {292}},\ \bibinfo {pages} {343} (\bibinfo {year}
  {2009})}\BibitemShut {NoStop}%
\bibitem [{\citenamefont {Bais}\ and\ \citenamefont
  {Bouwknegt}(1987)}]{ALEXANDERBAIS1987561}%
  \BibitemOpen
  \bibfield  {author} {\bibinfo {author} {\bibfnamefont {F.~A.}\ \bibnamefont
  {Bais}}\ and\ \bibinfo {author} {\bibfnamefont {P.~G.}\ \bibnamefont
  {Bouwknegt}},\ }\href {\doibase 10.1016/0550-3213(87)90010-1} {\bibfield
  {journal} {\bibinfo  {journal} {Nuclear Physics B}\ }\textbf {\bibinfo
  {volume} {279}},\ \bibinfo {pages} {561 } (\bibinfo {year}
  {1987})}\BibitemShut {NoStop}%
\bibitem [{\citenamefont {Garibaldi}(2016)}]{Garibaldi}%
  \BibitemOpen
  \bibfield  {author} {\bibinfo {author} {\bibfnamefont {S.}~\bibnamefont
  {Garibaldi}},\ }\href {\doibase 10.1090/bull/1540} {\bibfield  {journal}
  {\bibinfo  {journal} {Bulletin of the American Mathematical Society}\
  }\textbf {\bibinfo {volume} {53}},\ \bibinfo {pages} {643} (\bibinfo {year}
  {2016})}\BibitemShut {NoStop}%
\bibitem [{\citenamefont {Schellekens}(1993)}]{Schellekens}%
  \BibitemOpen
  \bibfield  {author} {\bibinfo {author} {\bibfnamefont {A.N.}~\bibnamefont
  {Schellekens}},\ }{\bibfield  {journal}
  {\bibinfo  {journal} {Communications in Mathematical Physics}\
  }\textbf {\bibinfo {volume} {153}},\ \bibinfo {pages} {159} (\bibinfo {year}
  {1993})}\BibitemShut {NoStop}%
\bibitem [{Note1()}]{Note1}%
  \BibitemOpen
  \bibinfo {note} {Technically, this follows trivially from the unimodularity of the $E_{8}$ Cartan matrix}\BibitemShut {NoStop}%
\bibitem [{\citenamefont {Di~Francesco}\ \emph {et~al.}(1999)\citenamefont
  {Di~Francesco}, \citenamefont {Mathieu},\ and\ \citenamefont
  {Senechal}}]{bigyellowbook}%
  \BibitemOpen
  \bibfield  {author} {\bibinfo {author} {\bibfnamefont {P.}~\bibnamefont
  {Di~Francesco}}, \bibinfo {author} {\bibfnamefont {P.}~\bibnamefont
  {Mathieu}}, \ and\ \bibinfo {author} {\bibfnamefont {D.}~\bibnamefont
  {Senechal}},\ }\href@noop {} {\emph {\bibinfo {title} {Conformal Field
  Theory}}}\ (\bibinfo  {publisher} {Springer, New York},\ \bibinfo {year}
  {1999})\BibitemShut {NoStop}%
\bibitem [{\citenamefont {Wess}\ and\ \citenamefont
  {Zumino}(1971)}]{WessZumino71}%
  \BibitemOpen
  \bibfield  {author} {\bibinfo {author} {\bibfnamefont {J.}~\bibnamefont
  {Wess}}\ and\ \bibinfo {author} {\bibfnamefont {B.}~\bibnamefont {Zumino}},\
  }\href {\doibase 10.1016/0370-2693(71)90582-X} {\bibfield  {journal}
  {\bibinfo  {journal} {Physics Letters B}\ }\textbf {\bibinfo {volume} {37}},\
  \bibinfo {pages} {95 } (\bibinfo {year} {1971})}\BibitemShut {NoStop}%
\bibitem [{\citenamefont {Witten}(1983)}]{WittenWZW}%
  \BibitemOpen
  \bibfield  {author} {\bibinfo {author} {\bibfnamefont {E.}~\bibnamefont
  {Witten}},\ }\href {\doibase 10.1016/0550-3213(83)90063-9} {\bibfield
  {journal} {\bibinfo  {journal} {Nuclear Physics B}\ }\textbf {\bibinfo
  {volume} {223}},\ \bibinfo {pages} {422 } (\bibinfo {year}
  {1983})}\BibitemShut {NoStop}%
\bibitem [{Note2()}]{Note2}%
  \BibitemOpen
  \bibinfo {note} {In fact, a solution exists for $N=9$ wires also, where the $E_{8}$
quantum Hall phase develops at filling fraction $\nu=32$, higher
than our present solution. Also, the Dynkin labels 4, 5, 6 and 8 in this construction are neutral, leaving an SO(8) subsector with trivial $\mathsf{x}$-momenta. A main consequence is that the embedding of $G_{2}$ currents also carry trivial momenta, and Fibonacci phases can never be stabilized.}\BibitemShut {Stop}%
\bibitem [{\citenamefont {Fradkin}(2013)}]{Fradkinbook}%
  \BibitemOpen
  \bibfield  {author} {\bibinfo {author} {\bibfnamefont {E.}~\bibnamefont
  {Fradkin}},\ }\href@noop {} {\emph {\bibinfo {title} {Field Theories of
  Condensed Matter Physics}}},\ \bibinfo {edition} {2nd}\ ed.\ (\bibinfo
  {publisher} {Cambridge University Press},\ \bibinfo {year}
  {2013})\BibitemShut {NoStop}%
\bibitem [{\citenamefont {Haldane}(1995)}]{Haldane95}%
  \BibitemOpen
  \bibfield  {author} {\bibinfo {author} {\bibfnamefont {F.~D.~M.}\
  \bibnamefont {Haldane}},\ }\href {\doibase 10.1103/PhysRevLett.74.2090}
  {\bibfield  {journal} {\bibinfo  {journal} {Phys. Rev. Lett.}\ }\textbf
  {\bibinfo {volume} {74}},\ \bibinfo {pages} {2090} (\bibinfo {year}
  {1995})}\BibitemShut {NoStop}%
\bibitem [{\citenamefont {Halperin}(1982)}]{Halperin82}%
  \BibitemOpen
  \bibfield  {author} {\bibinfo {author} {\bibfnamefont {B.~I.}\ \bibnamefont
  {Halperin}},\ }\href {\doibase 10.1103/PhysRevB.25.2185} {\bibfield
  {journal} {\bibinfo  {journal} {Phys. Rev. B}\ }\textbf {\bibinfo {volume}
  {25}},\ \bibinfo {pages} {2185} (\bibinfo {year} {1982})}\BibitemShut
  {NoStop}%
\bibitem [{\citenamefont {Wen}(1990)}]{WenchiralLL90}%
  \BibitemOpen
  \bibfield  {author} {\bibinfo {author} {\bibfnamefont {X.-G.}\ \bibnamefont
  {Wen}},\ }\href {\doibase 10.1103/PhysRevB.41.12838} {\bibfield  {journal}
  {\bibinfo  {journal} {Phys. Rev. B}\ }\textbf {\bibinfo {volume} {41}},\
  \bibinfo {pages}
  {\href{http://link.aps.org/doi/10.1103/PhysRevB.41.12838}{12838}} (\bibinfo
  {year} {1990})}\BibitemShut {NoStop}%
\bibitem [{\citenamefont {Kane}\ and\ \citenamefont
  {Fisher}(1997)}]{KaneFisher97}%
  \BibitemOpen
  \bibfield  {author} {\bibinfo {author} {\bibfnamefont {C.~L.}\ \bibnamefont
  {Kane}}\ and\ \bibinfo {author} {\bibfnamefont {M.~P.~A.}\ \bibnamefont
  {Fisher}},\ }\href {\doibase 10.1103/PhysRevB.55.15832} {\bibfield  {journal}
  {\bibinfo  {journal} {Phys. Rev. B}\ }\textbf {\bibinfo {volume} {55}},\
  \bibinfo {pages} {15832} (\bibinfo {year} {1997})}\BibitemShut {NoStop}%
\bibitem [{\citenamefont {Cappelli}\ \emph {et~al.}(2002)\citenamefont
  {Cappelli}, \citenamefont {Huerta},\ and\ \citenamefont
  {Zemba}}]{Cappelli01}%
  \BibitemOpen
  \bibfield  {author} {\bibinfo {author} {\bibfnamefont {A.}~\bibnamefont
  {Cappelli}}, \bibinfo {author} {\bibfnamefont {M.}~\bibnamefont {Huerta}}, \
  and\ \bibinfo {author} {\bibfnamefont {G.~R.}\ \bibnamefont {Zemba}},\ }\href
  {\doibase http://dx.doi.org/10.1016/S0550-3213(02)00340-1} {\bibfield
  {journal} {\bibinfo  {journal} {Nuclear Physics B}\ }\textbf {\bibinfo
  {volume} {636}},\ \bibinfo {pages} {568 } (\bibinfo {year}
  {2002})}\BibitemShut {NoStop}%
\bibitem [{\citenamefont {Vishwanath}\ and\ \citenamefont
  {Senthil}(2013)}]{VishwanathSenthil12}%
  \BibitemOpen
  \bibfield  {author} {\bibinfo {author} {\bibfnamefont {A.}~\bibnamefont
  {Vishwanath}}\ and\ \bibinfo {author} {\bibfnamefont {T.}~\bibnamefont
  {Senthil}},\ }\href {\doibase 10.1103/PhysRevX.3.011016} {\bibfield
  {journal} {\bibinfo  {journal} {Phys. Rev. X}\ }\textbf {\bibinfo {volume}
  {3}},\ \bibinfo {pages} {011016} (\bibinfo {year} {2013})}\BibitemShut
  {NoStop}%
\bibitem [{\citenamefont {Wang}\ \emph {et~al.}(2014)\citenamefont {Wang},
  \citenamefont {Potter},\ and\ \citenamefont {Senthil}}]{WangPotterSenthil13}%
  \BibitemOpen
  \bibfield  {author} {\bibinfo {author} {\bibfnamefont {C.}~\bibnamefont
  {Wang}}, \bibinfo {author} {\bibfnamefont {A.~C.}\ \bibnamefont {Potter}}, \
  and\ \bibinfo {author} {\bibfnamefont {T.}~\bibnamefont {Senthil}},\ }\href
  {\doibase 10.1126/science.1243326} {\bibfield  {journal} {\bibinfo  {journal}
  {Science}\ }\textbf {\bibinfo {volume} {343}},\ \bibinfo {pages} {629}
  (\bibinfo {year} {2014})}\BibitemShut {NoStop}%
\bibitem [{\citenamefont {Hasan}\ and\ \citenamefont
  {Kane}(2010)}]{HasanKane10}%
  \BibitemOpen
  \bibfield  {author} {\bibinfo {author} {\bibfnamefont {M.~Z.}\ \bibnamefont
  {Hasan}}\ and\ \bibinfo {author} {\bibfnamefont {C.~L.}\ \bibnamefont
  {Kane}},\ }\href {\doibase 10.1103/RevModPhys.82.3045} {\bibfield  {journal}
  {\bibinfo  {journal} {Rev. Mod. Phys.}\ }\textbf {\bibinfo {volume} {82}},\
  \bibinfo {pages} {3045} (\bibinfo {year} {2010})}\BibitemShut {NoStop}%
\bibitem [{\citenamefont {Qi}\ and\ \citenamefont
  {Zhang}(2011)}]{QiZhangreview11}%
  \BibitemOpen
  \bibfield  {author} {\bibinfo {author} {\bibfnamefont {X.-L.}\ \bibnamefont
  {Qi}}\ and\ \bibinfo {author} {\bibfnamefont {S.-C.}\ \bibnamefont {Zhang}},\
  }\href {\doibase 10.1103/RevModPhys.83.1057} {\bibfield  {journal} {\bibinfo
  {journal} {Rev. Mod. Phys.}\ }\textbf {\bibinfo {volume} {83}},\ \bibinfo
  {pages} {1057} (\bibinfo {year} {2011})}\BibitemShut {NoStop}%
\bibitem [{\citenamefont {Hasan}\ and\ \citenamefont
  {Moore}(2011)}]{HasanMoore11}%
  \BibitemOpen
  \bibfield  {author} {\bibinfo {author} {\bibfnamefont {M.~Z.}\ \bibnamefont
  {Hasan}}\ and\ \bibinfo {author} {\bibfnamefont {J.~E.}\ \bibnamefont
  {Moore}},\ }\href {https://doi.org/10.1146/annurev-conmatphys-062910-140432}
  {\bibfield  {journal} {\bibinfo  {journal} {Annual Review of Condensed Matter
  Physics}\ }\textbf {\bibinfo {volume} {2}},\ \bibinfo {pages} {55} (\bibinfo
  {year} {2011})}\BibitemShut {NoStop}%
\bibitem [{\citenamefont {Chiu}\ \emph {et~al.}(2016)\citenamefont {Chiu},
  \citenamefont {Teo}, \citenamefont {Schnyder},\ and\ \citenamefont
  {Ryu}}]{RMP}%
  \BibitemOpen
  \bibfield  {author} {\bibinfo {author} {\bibfnamefont {C.-K.}\ \bibnamefont
  {Chiu}}, \bibinfo {author} {\bibfnamefont {J.~C.~Y.}\ \bibnamefont {Teo}},
  \bibinfo {author} {\bibfnamefont {A.~P.}\ \bibnamefont {Schnyder}}, \ and\
  \bibinfo {author} {\bibfnamefont {S.}~\bibnamefont {Ryu}},\ }\href {\doibase
  10.1103/RevModPhys.88.035005} {\bibfield  {journal} {\bibinfo  {journal}
  {Rev. Mod. Phys.}\ }\textbf {\bibinfo {volume} {88}},\ \bibinfo {pages}
  {035005} (\bibinfo {year} {2016})}\BibitemShut {NoStop}%
\bibitem [{\citenamefont {Cano}\ \emph {et~al.}(2014)\citenamefont {Cano},
  \citenamefont {Cheng}, \citenamefont {Mulligan}, \citenamefont {Nayak},
  \citenamefont {Plamadeala},\ and\ \citenamefont {Yard}}]{cano2013bulk}%
  \BibitemOpen
  \bibfield  {author} {\bibinfo {author} {\bibfnamefont {J.}~\bibnamefont
  {Cano}}, \bibinfo {author} {\bibfnamefont {M.}~\bibnamefont {Cheng}},
  \bibinfo {author} {\bibfnamefont {M.}~\bibnamefont {Mulligan}}, \bibinfo
  {author} {\bibfnamefont {C.}~\bibnamefont {Nayak}}, \bibinfo {author}
  {\bibfnamefont {E.}~\bibnamefont {Plamadeala}}, \ and\ \bibinfo {author}
  {\bibfnamefont {J.}~\bibnamefont {Yard}},\ }\href {\doibase
  10.1103/PhysRevB.89.115116} {\bibfield  {journal} {\bibinfo  {journal} {Phys.
  Rev. B}\ }\textbf {\bibinfo {volume} {89}},\ \bibinfo {pages} {115116}
  (\bibinfo {year} {2014})}\BibitemShut {NoStop}%
\bibitem [{\citenamefont {{Sirota}}\ \emph {et~al.}(2018)\citenamefont
  {{Sirota}}, \citenamefont {{Sahoo}}, \citenamefont {{Cho}},\ and\
  \citenamefont {{Teo}}}]{SirotaSahooChoTeo18}%
  \BibitemOpen
  \bibfield  {author} {\bibinfo {author} {\bibfnamefont {A.}~\bibnamefont
  {{Sirota}}}, \bibinfo {author} {\bibfnamefont {S.}~\bibnamefont {{Sahoo}}},
  \bibinfo {author} {\bibfnamefont {G.~Y.}\ \bibnamefont {{Cho}}}, \ and\
  \bibinfo {author} {\bibfnamefont {J.~C.~Y.}\ \bibnamefont {{Teo}}},\
  }\href@noop {} {\bibfield  {journal} {\bibinfo  {journal} {arXiv e-prints}\
  ,\ \bibinfo {eid} {arXiv:1812.01642}} (\bibinfo {year} {2018})},\ \Eprint
  {http://arxiv.org/abs/1812.01642} {arXiv:1812.01642 [cond-mat.str-el]}
  \BibitemShut {NoStop}%
\bibitem [{\citenamefont {Mross}\ \emph
  {et~al.}(2016{\natexlab{a}})\citenamefont {Mross}, \citenamefont {Alicea},\
  and\ \citenamefont {Motrunich}}]{MrossAliceaMotrunich16}%
  \BibitemOpen
  \bibfield  {author} {\bibinfo {author} {\bibfnamefont {D.~F.}\ \bibnamefont
  {Mross}}, \bibinfo {author} {\bibfnamefont {J.}~\bibnamefont {Alicea}}, \
  and\ \bibinfo {author} {\bibfnamefont {O.~I.}\ \bibnamefont {Motrunich}},\
  }\href {\doibase 10.1103/PhysRevLett.117.016802} {\bibfield  {journal}
  {\bibinfo  {journal} {Phys. Rev. Lett.}\ }\textbf {\bibinfo {volume} {117}},\
  \bibinfo {pages} {016802} (\bibinfo {year} {2016}{\natexlab{a}})}\BibitemShut
  {NoStop}%
\bibitem [{\citenamefont {Mross}\ \emph
  {et~al.}(2016{\natexlab{b}})\citenamefont {Mross}, \citenamefont {Alicea},\
  and\ \citenamefont {Motrunich}}]{MrossAliceaMotrunich16PRL}%
  \BibitemOpen
  \bibfield  {author} {\bibinfo {author} {\bibfnamefont {D.~F.}\ \bibnamefont
  {Mross}}, \bibinfo {author} {\bibfnamefont {J.}~\bibnamefont {Alicea}}, \
  and\ \bibinfo {author} {\bibfnamefont {O.~I.}\ \bibnamefont {Motrunich}},\
  }\href {\doibase 10.1103/PhysRevLett.117.136802} {\bibfield  {journal}
  {\bibinfo  {journal} {Phys. Rev. Lett.}\ }\textbf {\bibinfo {volume} {117}},\
  \bibinfo {pages} {136802} (\bibinfo {year} {2016}{\natexlab{b}})}\BibitemShut
  {NoStop}%
\bibitem [{\citenamefont {Mross}\ \emph {et~al.}(2017)\citenamefont {Mross},
  \citenamefont {Alicea},\ and\ \citenamefont
  {Motrunich}}]{MrossAliceaMotrunich17}%
  \BibitemOpen
  \bibfield  {author} {\bibinfo {author} {\bibfnamefont {D.~F.}\ \bibnamefont
  {Mross}}, \bibinfo {author} {\bibfnamefont {J.}~\bibnamefont {Alicea}}, \
  and\ \bibinfo {author} {\bibfnamefont {O.~I.}\ \bibnamefont {Motrunich}},\
  }\href {\doibase 10.1103/PhysRevX.7.041016} {\bibfield  {journal} {\bibinfo
  {journal} {Phys. Rev. X}\ }\textbf {\bibinfo {volume} {7}},\ \bibinfo {pages}
  {041016} (\bibinfo {year} {2017})}\BibitemShut {NoStop}%
\bibitem [{\citenamefont {{Fuji}}\ and\ \citenamefont
  {{Furusaki}}(2018)}]{FujiFurusaki18}%
  \BibitemOpen
  \bibfield  {author} {\bibinfo {author} {\bibfnamefont {Y.}~\bibnamefont
  {{Fuji}}}\ and\ \bibinfo {author} {\bibfnamefont {A.}~\bibnamefont
  {{Furusaki}}},\ }\href@noop {} {\bibfield  {journal} {\bibinfo  {journal}
  {ArXiv e-prints}\ } (\bibinfo {year} {2018})},\ \Eprint
  {http://arxiv.org/abs/1808.07648} {arXiv:1808.07648 [cond-mat.str-el]}
  \BibitemShut {NoStop}%
\bibitem [{\citenamefont {{Banerjee}}\ \emph {et~al.}(2017)\citenamefont
  {{Banerjee}}, \citenamefont {{Heiblum}}, \citenamefont {{Rosenblatt}},
  \citenamefont {{Oreg}}, \citenamefont {{Feldman}}, \citenamefont {{Stern}},\
  and\ \citenamefont
  {{Umansky}}}]{BanerjeeHeiblumRosenblattOregFeldmanSternUmansky17}%
  \BibitemOpen
  \bibfield  {author} {\bibinfo {author} {\bibfnamefont {M.}~\bibnamefont
  {{Banerjee}}}, \bibinfo {author} {\bibfnamefont {M.}~\bibnamefont
  {{Heiblum}}}, \bibinfo {author} {\bibfnamefont {A.}~\bibnamefont
  {{Rosenblatt}}}, \bibinfo {author} {\bibfnamefont {Y.}~\bibnamefont
  {{Oreg}}}, \bibinfo {author} {\bibfnamefont {D.~E.}\ \bibnamefont
  {{Feldman}}}, \bibinfo {author} {\bibfnamefont {A.}~\bibnamefont {{Stern}}},
  \ and\ \bibinfo {author} {\bibfnamefont {V.}~\bibnamefont {{Umansky}}},\
  }\href {\doibase 10.1038/nature22052} {\bibfield  {journal} {\bibinfo
  {journal} {Nature}\ }\textbf {\bibinfo {volume} {545}},\ \bibinfo {pages}
  {75} (\bibinfo {year} {2017})}\BibitemShut {NoStop}%
\bibitem [{\citenamefont {{Banerjee}}\ \emph {et~al.}(2018)\citenamefont
  {{Banerjee}}, \citenamefont {{Heiblum}}, \citenamefont {{Umansky}},
  \citenamefont {{Feldman}}, \citenamefont {{Oreg}},\ and\ \citenamefont
  {{Stern}}}]{BanerjeeHeiblumUmanskyFeldmanOregStern18}%
  \BibitemOpen
  \bibfield  {author} {\bibinfo {author} {\bibfnamefont {M.}~\bibnamefont
  {{Banerjee}}}, \bibinfo {author} {\bibfnamefont {M.}~\bibnamefont
  {{Heiblum}}}, \bibinfo {author} {\bibfnamefont {V.}~\bibnamefont
  {{Umansky}}}, \bibinfo {author} {\bibfnamefont {D.~E.}\ \bibnamefont
  {{Feldman}}}, \bibinfo {author} {\bibfnamefont {Y.}~\bibnamefont {{Oreg}}}, \
  and\ \bibinfo {author} {\bibfnamefont {A.}~\bibnamefont {{Stern}}},\ }\href
  {\doibase 10.1038/s41586-018-0184-1} {\bibfield  {journal} {\bibinfo
  {journal} {Nature}\ }\textbf {\bibinfo {volume} {559}},\ \bibinfo {pages}
  {205} (\bibinfo {year} {2018})}\BibitemShut {NoStop}%
\bibitem [{\citenamefont {Macfarlane}(2001)}]{MacfarlaneG2}%
  \BibitemOpen
  \bibfield  {author} {\bibinfo {author} {\bibfnamefont {A.~J.}\ \bibnamefont
  {Macfarlane}},\ }\href {\doibase 10.1142/S0217751X01004335} {\bibfield
  {journal} {\bibinfo  {journal} {International Journal of Modern Physics A}\
  }\textbf {\bibinfo {volume} {16}},\ \bibinfo {pages} {3067} (\bibinfo {year}
  {2001})}\BibitemShut {NoStop}%

\end{thebibliography}

\newpage

\appendix

\section{\texorpdfstring{$E_{8}$}{E8} Quantum Hall state momentum conservation ~\label{subsec:App_momenta}}

We present here the solution of the momentum commensurability conditions stated in the main text, Eq.~\eqref{eq:E8momenta_fix}. There are 11 vanishing (mod $2\pi$)
linear equations for the 11 unknown momenta, with coefficients that are also linear in the inverse of the filling fraction $\nu$. A non-trivial solution to $k_{F,a}$ exists only for a vanishing determinant which fixes $\nu$ as
\begin{equation}
\frac{\nu-16}{\nu}=0\implies\nu=16.
\end{equation}
Plugging back $\nu=16$ into Eq.~\eqref{eq:E8momenta_fix}  and solving for the momenta returns

\begin{gather}
k_{y,1}^{\sigma}=k_{y,2}^{\sigma}=\frac{1}{2}yk_{F},\;k_{y,3}^{\sigma}=k_{y,7}^{\sigma}=\frac{1}{2}\left(y-\sigma\right)k_{F},\\
k_{y,4}^{\sigma}=k_{y,5}^{\sigma}=k_{y,6}^{\sigma}=k_{y,8}^{\sigma}=k_{y,9}^{\sigma}=\frac{1}{2}\left(y+2\sigma\right)k_{F},\\
k_{y,10}^{\sigma}=\frac{1}{2}\left(y+3\sigma\right)k_{F},\;k_{y,11}^{\sigma}=\frac{1}{2}\left(y-3\sigma\right)k_{F}.
\end{gather}
With these, the $\sigma=L$ and $R$ channels of any of the three
recombined fermions $f_{yn}^{\sigma}$, for $n=1,2,3$, share the
same momentum, and therefore the oscillatory terms in the intra-bundle
backscattering interactions of Eq.~(\ref{eq:IntraU1}) cancel. Similarly,
the inter-bundle terms in (\ref{eq:InterE8}) also conserve momentum,
as $\tilde{k}_{y,I}^{R}=\tilde{k}_{y+1,I}^{L}$ for $I=1,\ldots,8$.

\section{A \texorpdfstring{$G_{2}\times F_{4}$}{G2,F4} energy-momentum tensor
~\label{subsec:G2F4_EM_tensor}}

Here we compare the energy-momentum tensors of the $E_8$, $G_2$ and $F_4$ theories at level 1. The goal is to see that, through our embedding, an exact decomposition of the operators is obtained. By definition, \hyperlink{WZW}{WZW} energy-momentum tensors at level 1 read~\cite{bigyellowbook}

\begin{equation}
T\left(z\right)=\frac{\left(\mathbf{J}\cdot\mathbf{J}\right)\left(z\right)}{2\left(1+g\right)},
\end{equation}
with $J^{a}$ the Sugawara current, $g$ dual coxeter number, and the
normal ordering defined as

\begin{equation}
\left(J^{a}J^{a}\right)\left(z\right)=\frac{1}{2\pi i}\oint_{z}\frac{dw}{w-z}J^{a}\left(w\right)J^{a}\left(z\right).
\end{equation}
The contraction of the Sugawara currents can be written in the Cartan-Weyl
basis

\begin{equation}
\left(\mathbf{J}\cdot\mathbf{J}\right)\left(z\right)=\sum_{j}\left(H^{j}H^{j}\right)\left(z\right)+\sum_{\boldsymbol{\alpha}}\left(E^{-\boldsymbol{\alpha}}E^{\boldsymbol{\alpha}}\right)\left(z\right),
\end{equation}
where the $\boldsymbol{\alpha}$ sum is over the full root lattice 
while $j$ sums over the generators of the Cartan subalgebra. We have
absorbed the normalization factors into the root operators.

We are then ready to verify the energy-momentum tensor decoupling
via the conformal embedding. Under the $SO(16)$ embedding, the $E_{8}$
tensor reduces to 
\begin{align}
T_{E_{8}}\left(z\right) & =-\frac{\partial\boldsymbol{\phi}\cdot\partial\boldsymbol{\phi}}{2},
\end{align}
which is, in fact, of the same form of the $SO(16)$ energy-momentum
tensor.

To fully verify the conformal embedding, one may compute the energy
momentum tensors of the $G_{2}$ and $F_{4}$ \hyperlink{CFT}{CFT}s. This calculation
requires lengthy but straightforward bookkeeping, and will not be
presented in here. The operators $T_{G_{2}}$ and $T_{F_{4}}$ are
found to be
\begin{widetext}
\begin{align}
T_{G_{2}}\left(z\right) & =-\frac{1}{2}\left[\left(\sum_{j=1}^{3}\partial\phi_{j}\partial\phi_{j}\right)\left(z\right)-\frac{1}{5}\left(\sum_{j=1}^{3}\partial\phi_{j}\right)^{2}\left(z\right)\right]-\frac{1}{5}\left(\partial\phi_{4}\partial\phi_{4}\right)\left(z\right)\nonumber \\
 & +\frac{2}{5}\left\{ \cos\left[2\left(\frac{\pi}{8}-\phi_{+}\left(z\right)\right)\right]-\cos\left[2\left(\frac{\pi}{8}-\phi_{-}\left(z\right)\right)\right]+\cos\left[2\phi_{4}\left(z\right)\right]\right\} ,
\end{align}
and 
\begin{align}
T_{F_{4}}\left(z\right)= & -\frac{1}{2}\left[\sum_{j=5}^{8}\left(\partial\phi_{j}\partial\phi_{j}\right)\left(z\right)+\frac{1}{5}\left(\sum_{j=1}^{3}\partial\phi_{j}\right)^{2}\left(z\right)\right]-\frac{3}{10}\left(\partial\phi_{4}\partial\phi_{4}\right)\left(z\right)\nonumber \\
 & -\frac{2}{5}\left\{ \cos\left[2\left(\frac{\pi}{8}-\phi_{+}\left(z\right)\right)\right]-\cos\left[2\left(\frac{\pi}{8}-\phi_{-}\left(z\right)\right)\right]+\cos\left[2\phi_{4}\left(z\right)\right]\right\} ,
\end{align}
\end{widetext}

where $\phi_{\pm}\equiv\phi_{1}+\phi_{2}+\phi_{3}\pm\phi_{4}$. The
sum of these two expressions returns $T_{E_{8}}$, as it should, finishing
the verification of the conformal embedding. 

\section{\texorpdfstring{$G_{2}$}{G2}
and \texorpdfstring{$F_{4}$}{F4} quantum Hall states momentum commensurability conditions~\label{subsec:Fermi-momenta-and-filling-G2-F4}}

To stabilize the $G_{2}$ and $F_{4}$ Fibonacci phases, a process
of fixing a distribution of Fermi momenta for the 11 electronic channels appearing in Eq.~(\ref{elecboson}) is necessary. This procedure is analogous to the one used for the $E_{8}$ Quantum Hall state in Appendix~\ref{subsec:App_momenta}. Demanding commensurability
conditions on the momenta for the $F_{4}$ Fibonacci phase in Eq.~(\ref{eq:F4back}) so that oscillatory terms cancel results in the unique non-trivial solution
(up to the single free parameter $k_{F}$) yields
\begin{gather}
k_{y,1}^{\sigma}=k_{y,2}^{\sigma}=k_{y,7}^{\sigma}=\left(y-\sigma\right)k_{F},\;k_{y,3}^{\sigma}=\left(y-2\sigma\right)k_{F},\nonumber \\
k_{y,4}^{\sigma}=k_{y,5}^{\sigma}=k_{y,6}^{\sigma}=k_{y,8}^{\sigma}=k_{y,9}^{\sigma}=\left(y+2\sigma\right)k_{F},\nonumber \\
k_{y,10}^{\sigma}=\left(y+3\sigma\right)k_{F},\;k_{y,11}^{\sigma}=\left(y-4\sigma\right)k_{F},\;\nu=8.
\end{gather}

Similarly, demanding momentum commensurability in Eq.~(\ref{eq:G2back}),
one obtains the Fermi momentum distribution for the coupled wire model
for the $G_{2}$ Fibonacci quantum Hall state, 
\begin{gather}
k_{y,1}^{\sigma}=k_{y,2}^{\sigma}=k_{y,3}^{\sigma}=k_{y,11}^{\sigma}=\left(\sigma+y\right)k_{F},\nonumber \\
k_{y,4}^{\sigma}=k_{y,5}^{\sigma}=k_{y,6}^{\sigma}=k_{y,7}^{\sigma}=k_{y,8}^{\sigma}=k_{y,9}^{\sigma}=k_{y,10}^{\sigma}=yk_{F},\nonumber \\\nu=8.
\end{gather}

\section{Fibonacci primary field representations in the \texorpdfstring{$G_{2}$}{G2} and \texorpdfstring{$F_{4}$}{F4} WZW CFTs at level 1}\label{subsec:Fibonacci-primary-fields}

Our prime motivation for studying $(G_{2})_{1}$ and $(F_{4})_{1}$
\hyperlink{WZW}{WZW} theories stems from the claim that both carry excitations in the
form of Fibonacci anyons. Here we will provide a short demonstration
of that, and then follow with a coset construction that allows us
to profit from the embeddings discussed up to now to explicitly build
the corresponding Fibonacci primary fields.

To see that the only excitations in $(G_{2})_{1}$ and $(F_{4})_{1}$
are Fibonacci anyons, we can start by noticing that at level 1, these
theories contain only one non-trivial primary field besides the vacuum
$\mathbb{I}$. We name these fields $\tau$ for $(G_{2})_{1}$ and
$\bar{\tau}$ for $(F_{4})_{1}$. Following, we invoke the Gauss-Milgram
formula; this formula is a manifestation of the bulk-boundary correspondence as it connects quantities that point to the bulk anyon excitations of
a topological phase to the \hyperlink{CFT}{CFT} degrees of freedom that live at its
boundary. Stating the formula explicitly,
\begin{equation}
\sum_{a}d_{a}^{2}\theta_{a}=\mathcal{D}e^{i2\pi\frac{c}{8}},\label{eq:Gauss-Milgram}
\end{equation}
where $\mathcal{D}^{2}\equiv\sum_{a}d_{a}^{2}$ is the total quantum
order expressed in terms of the quantum dimensions $d_{a}$, quantities that characterize the bulk anyons. The conformal spins are $\theta_{a}=e^{i2\pi h_{a}}$, determined by quantum dimensions $h_{a}$, and $c$
is the chiral central charge. The latter two quantities characterize
the \hyperlink{CFT}{CFT} at the edge of the topological phase. The sum is over all primary fields of the \hyperlink{CFT}{CFT} or, correspondingly, all anyons. 

The conformal dimension of a primary field $a$ of a \hyperlink{WZW}{WZW} theory is completely determined by its Lie algebra content by $h_{a}=\frac{C_{a}}{2\left(k+g\right)},$~\citep{bigyellowbook}
where $k$ is the level, $g$ is the dual coxeter number and $C_{a}$
is the quadratic Casimir of the representation. Let us consider a simple example first: trivial topological order. In this case we just have the trivial identity anyon $a=1$ and $\mathcal{D}=1$. The Gauss-Milgram formula returns $e^{i2\pi\frac{c}{8}}=1$, enforcing that trivial anyon statistics implies that the central charge is defined only modulo 8, as discussed at the introduction.

Moving forward, we consider the $G_2$ and $F_4$ cases. Collecting the dual Coxeter number and the quadratic Casimir, we obtain, $h_{\tau}=2/5$ and $h_{\bar{\tau}}=3/5$. Furthermore, $d_{\mathbb{I}}=1$ and $h_{\mathbb{I}}=0$, leaving a single unknown
in the Gauss-Milgram formula \eqref{eq:Gauss-Milgram}, namely $d_{\tau}$ or $d_{\bar{\tau}}$
for $G_{2}$ or $F_{4}$. Solving for these, 

\begin{equation}
d_{\tau}=d_{\bar{\tau}}=\frac{1+\sqrt{5}}{2},
\end{equation}
which is the Golden ratio expected for Fibonacci anyons. Since the
quantum dimensions obey a algebraic version of the fusion rules,
these follow imediately as $\tau\times\tau=\mathbb{I}+\tau$. Equivalently,
the fusion rules can be explicitly determined by the modular ($2\times2$)
S-matrices of the theory using the Verlinde formula\cite{bigyellowbook}.

We thus established that the chiral $(G_{2})_{1}$ and $(F_{4})_{1}$
\hyperlink{WZW}{WZW} edge \hyperlink{CFT}{CFT}s contain primary fields that obey the Fibonacci fusion
rules. They correspond to Fibonacci anyonic excitations in the 2D
bulk, and thus we refer to them as Fibonacci primary fields. Let us
now construct explicit expressions for them based on our conformal
embedding here developed.

The non-trivial primary fields $[\tau]$ and $[\bar{\tau}]$ are associated
with the fundamental irreducible representations of their respective
exceptional Lie algebras. Each of them consists of a super-selection
sector of fields, $[\tau]=\mathrm{span}\{\tau_{m}\}_{m=1,\ldots,7}$
and $[\bar{\tau}]=\mathrm{span}\{\bar{\tau}_{l}\}_{l=1,\ldots,26}$,
that rotate into each other by the \hyperlink{WZW}{WZW} algebraic actions 
\begin{align}
\left[E_{G_{2}}(\mathsf{z})\right]_{\boldsymbol{\gamma}}\tau_{m}(\mathsf{w})=\frac{1}{\mathsf{z}-\mathsf{w}}\rho_{G_{2}}(\boldsymbol{\gamma})_{m}^{m'}\tau_{m'}(\mathsf{w})+\ldots,\nonumber\\\quad\left[E_{F_{4}}(\mathsf{z})\right]_{\boldsymbol{\beta}}\bar{\tau}_{l}(\mathsf{w})=\frac{1}{\mathsf{z}-\mathsf{w}}\rho_{F_{4}}(\boldsymbol{\beta})_{l}^{l'}\bar{\tau}_{l'}(\mathsf{w})+\ldots,
\end{align}
where $\mathsf{z},\mathsf{w}\sim e^{\tau+i\mathsf{x}}$ are radially
ordered holomorphic space-time parameters, $\boldsymbol{\gamma}$
and $\boldsymbol{\beta}$ are the roots of $G_{2}$ and $F_{4}$,
and $\rho_{G_{2}}$ and $\rho_{F_{4}}$ are the 7- and 26-dimensional
irreducible matrix representation of the $G_{2}$ and $F_{4}$ algebras.
Here, we provide parafermionic representations of these fields that
constitute the Fibonacci super-sectors. Using the coset construction,
each Fibonacci field $\tau_{m}$, $\bar{\tau}_{l}$ can be expressed
as a product of two components: (1) a non-Abelian primary field of
the $\mathbb{Z}_{3}$ parafermion \hyperlink{CFT}{CFT} or the tricritical Ising \hyperlink{CFT}{CFT},
respectively, and (2) a vertex operator of bosonized variables.

The $(G_{2})_{1}$ \hyperlink{WZW}{WZW} \hyperlink{CFT}{CFT} can be decomposed into two decoupled sectors
using its $SU(3)_{1}$ sub-algebra. 
\begin{align}
(G_{2})_{1}\simeq SU(3)_{1}\times\frac{(G_{2})_{1}}{SU(3)_{1}}=SU(3)_{1}\times\mbox{\ensuremath{\mathbb{Z}_{3}} parafermion}.
\end{align}
For instance, the decomposition agrees with the partition of the energy-momentum
tensors $T_{\mathbb{Z}_{3}}=T_{(G_{2})_{1}/SU(3)_{1}}\equiv T_{(G_{2})_{1}}-T_{SU(3)_{1}}$
and central charges $c((G_{2})_{1})=14/5=c(SU(3)_{1})+c(\mathbb{Z}_{3})=2+4/5$.
First, we focus on the $SU(3)_{1}$ sub-algebra. Using the aforementioned
fermionization of $E_{8}$, the six roots of $SU(3)$ coincide with
the long roots of $G_{2}$, $e^{\pm i(\phi_{1}-\phi_{2})}$, $e^{\pm i(\phi_{2}-\phi_{3})}$,
$e^{\pm i(\phi_{1}-\phi_{3})}$. The $SU(3)_{1}$ \hyperlink{WZW}{WZW} sub-algebra
has three primary fields, $\mathbb{I}$, $[\mathcal{E}]$ and $[\mathcal{E}^{-1}]$,
with conformal dimensions $h_{\mathbb{I}}=0$ and $h_{\mathcal{E}}=h_{\mathcal{E}^{-1}}=1/3$.
$\mathbb{I}$ denotes the trivial vacuum, while $[\mathcal{E}]$ and
$[\mathcal{E}^{-1}]$ are three-dimensional super-selection sectors
of fields
\begin{widetext} 
\begin{align}
\begin{split}[\mathcal{E}] & =\mathrm{span}\left\{ e^{i(\phi_{1}+\phi_{2}-2\phi_{3})/3},e^{i(\phi_{2}+\phi_{3}-2\phi_{1})/3},e^{i(\phi_{3}+\phi_{1}-2\phi_{2})/3}\right\} ,\\{}
[\mathcal{E}^{-1}] & =\mathrm{span}\left\{ e^{-i(\phi_{1}+\phi_{2}-2\phi_{3})/3},e^{-i(\phi_{2}+\phi_{3}-2\phi_{1})/3},e^{-i(\phi_{3}+\phi_{1}-2\phi_{2})/3}\right\} ,
\end{split}
\end{align}
\end{widetext}
that rotate according to the two fundamental representations of $SU(3)$.
For example, under the $SU(3)_{1}$ roots, 
\begin{align}
e^{i[\phi_{a}(\mathsf{z})-\phi_{b}(\mathsf{z})]}e^{i[\phi_{b}(\mathsf{w})+\phi_{c}(\mathsf{w})-2\phi_{a}(\mathsf{w})]}\sim \nonumber \\e^{i[\phi_{a}(\mathsf{w})+\phi_{c}(\mathsf{w})-2\phi_{b}(\mathsf{w})]}/(\mathsf{z}-\mathsf{w})+\ldots.
\end{align}
The 7-dimensional fundamental representation of $G_{2}$ decomposes
into $1+3+3$ under $SU(3)$ and each component is associated to a
distinct $SU(3)_{1}$ primary field.

Next, we focus on the $(G_{2})_{1}/SU(3)_{1}$ coset, which is identical
to the $\mathbb{Z}_{3}$ parafermionic \hyperlink{CFT}{CFT}. It supports three Abelian
primary fields $\mathbb{I},\Psi,\Psi^{-1}$ and three non-Abelian
ones $\tau,\varepsilon,\varepsilon^{-1}$. They have conformal dimensions
$h_{\mathbb{I}}=0$, $h_{\Psi}=h_{\Psi^{-1}}=2/3$, $h_{\tau}=2/5$
and $h_{\varepsilon}=h_{\varepsilon^{-1}}=1/15$. They obey the fusion
rules 
\begin{gather}
\Psi\times\Psi=\Psi^{-1},\;\Psi\times\Psi^{-1}=\mathbb{I},\;\tau\times\Psi=\varepsilon,\;\nonumber\\\tau\times\Psi^{-1}=\varepsilon^{-1},\;\tau\times\tau=\mathbb{I}+\tau.
\end{gather}
The Fibonacci primary field of $(G_{2})_{1}$ is the 7-dimensional
super-selection sector 
\begin{align}
[\tau] & =(\tau\otimes\mathbb{I})\oplus(\varepsilon\otimes[\mathcal{E}])\oplus(\varepsilon^{-1}\otimes[\mathcal{E}^{-1}])\nonumber \\
 & =\mathrm{span}\left\{ \begin{array}{c}
\tau,\varepsilon e^{i(\phi_{1}+\phi_{2}-2\phi_{3})/3},\\\varepsilon e^{i(\phi_{2}+\phi_{3}-2\phi_{1})/3},\varepsilon e^{i(\phi_{3}+\phi_{1}-2\phi_{2})/3},\\
\varepsilon^{-1}e^{-i(\phi_{1}+\phi_{2}-2\phi_{3})/3},\varepsilon^{-1}e^{-i(\phi_{2}+\phi_{3}-2\phi_{1})/3},\\\varepsilon^{-1}e^{-i(\phi_{3}+\phi_{1}-2\phi_{2})/3}
\end{array}\right\} 
\end{align}
All seven fields share the same conformal dimension $h_{\tau}=2/5$.
For example, $h_{\varepsilon\otimes[\mathcal{E}]}=1/15+1/3=2/5$.
The super-sector splits into three components under $SU(3)$. However,
they rotate irreducibly into each other under $G_{2}$.

The Fibonacci primary field of $(F_{4})_{1}$ can be described in
a similar manner. First, using the $SO(9)_{1}$ sub-algebra, the \hyperlink{WZW}{WZW}
\hyperlink{CFT}{CFT} can be factored into two decoupled sectors 
\begin{align}
(F_{4})_{1}\simeq SO(9)_{1}\times\frac{(F_{4})_{1}}{SO(9)_{1}}=SO(9)_{1}\times(\mbox{tricritical Ising}).
\end{align}
Like the previous $G_{2}$ coset decomposition, here the energy-momentum
tensor and central charge also decompose accordingly: $c((F_{4})_{1})=26/5=9/2+7/10$,
where $9/2$ and $7/10$ are the central charges for $SO(9)_{1}$
and the tricritical Ising \hyperlink{CFT}{CFT}s. The Fibonacci super-selection sector
of $(F_{4})_{1}$ consists of fields, which are linear combinations
of products of primary fields in $SO(9)_{1}$ and the tricritical
Ising \hyperlink{CFT}{CFT}s.

We first concentrate on $SO(9)_{1}$. It supports three primary fields
$\mathbb{I}$, $[\psi]$ and $[\Sigma]$ with conformal dimensions
$h_{\mathbb{I}}=0$, $h_{\psi}=1/2$ and $h_{\Sigma}=9/16$ and respectively
associate to the trivial, vector and spinor representations of $SO(9)$.
Using the fermionization convention of $E_{8}$, the $SO(9)_{1}$
theory is generated by the 9 Majorana fermions $\psi_{8},\ldots,\psi_{16}$,
where the last 8 Majorana fermions are paired into the 4 Dirac fermions
$d_{I}=(\psi_{2I-1}+i\psi_{2I})/\sqrt{2}\sim e^{i\phi_{I}}$, for
$I=5,6,7,8$. The vector primary field consists of any linear combinations
of these 9 fermions $[\psi]=\mathrm{span}\{\psi_{8},\ldots,\psi_{16}\}$.
We arbitrarily single out the first Majorana fermion $\psi_{8}$,
which is not paired with any of the others, and associate it to an
Ising \hyperlink{CFT}{CFT}. This further decomposes 
\begin{align}
SO(9)_{1}=\mathrm{Ising}\times SO(8)_{1}.
\end{align}
The spinor primary field of $SO(9)_{1}$ decomposes into a product
between the Ising twist field $\sigma$ and the $SO(8)_{1}$ spinors.
\begin{align}
[\Sigma]=\mathrm{span}\left\{ \sigma\exp\left(\frac{i}{2}\sum_{I=5}^{8}\epsilon^{I}\phi_{I}\right):\epsilon^{5},\ldots,\epsilon^{8}=\pm1\right\} .
\end{align}
The conformal dimension of $\sigma$ is $1/16$ and that of the $SO(8)_{1}$
spinors are $1/2$. Thus, they combine to the appropriate conformal
dimension of $h_{\Sigma}=9/16$ for each field in the set. The dimension
of the $SO(9)$ spinor representation is $2^{4}=16$. The 26-dimensional
fundamental representation of $F_{4}$ decomposes into $1+9+16$ under
the $SO(9)$ sub-algebra, and each component is associated to a unique
$SO(9)_{1}$ primary field.

We now focus on the $(F_{4})_{1}/SO(9)_{1}$ coset, which is identical
to the tricritical Ising \hyperlink{CFT}{CFT}, or equivalently, the minimal theory
$\mathcal{M}(5,4)$. The theory has six primary fields arranged in
the following conformal grid
\begin{align}
\begin{array}{|c|c|c|}
\hline f & s & \mathbb{I}\\
\hline \bar{\tau} & s\bar{\tau} & f\bar{\tau}\\
\hline f\bar{\tau} & s\bar{\tau} & \bar{\tau}\\
\hline \mathbb{I} & s & f
\\\hline \end{array}=\begin{array}{|c|c|c|}
\hline \Phi_{3,1} & \Phi_{2,1} & \Phi_{1,1}\\
\hline \Phi_{3,2} & \Phi_{2,2} & \Phi_{1,2}\\
\hline \Phi_{1,2} & \Phi_{2,2} & \Phi_{3,2}\\
\hline \Phi_{1,1} & \Phi_{2,1} & \Phi_{3,1}
\\\hline \end{array}\xrightarrow{\mbox{\scriptsize c.d.}}\begin{array}{|c|c|c|}
\hline 3/2 & 7/16 & 0\\
\hline 3/5 & 3/80 & 1/10\\
\hline 1/10 & 3/80 & 3/5\\
\hline 0 & 7/16 & 3/2
\\\hline \end{array}
\end{align}
with c.d. standing for conformal dimension.
They obey the fusion rules 
\begin{align}
f\times f=\mathbb{I},\;s\times f=s,\;s\times s=1+f,f\times\bar{\tau}=f\bar{\tau},\;\nonumber\\s\times\bar{\tau}=s\bar{\tau},\;\bar{\tau}\times\bar{\tau}=\mathbb{I}+\bar{\tau}.
\end{align}
The Fibonacci primary field of $(F_{4})_{1}$ is the 26-dimensional
super-selection sector 
\begin{align}
[\bar{\tau}] & =(\bar{\tau}\otimes\mathbb{I})\oplus(f\bar{\tau}\otimes[\psi])\oplus(s\bar{\tau}\otimes[\Sigma])\nonumber \\
 & =\mathrm{span}\left\{ \bar{\tau},f\bar{\tau}\psi_{j},s\bar{\tau}\sigma\exp\left(\frac{i}{2}\sum_{I=5}^{8}\epsilon^{I}\phi_{I}\right):\begin{array}{l}
j=8,\ldots,16\\
\epsilon^{5},\ldots,\epsilon^{8}=\pm1
\end{array}\right\} .
\end{align}
Each of these fields carry the identical conformal dimension $h_{\bar{\tau}}=3/5$.
For example, the second field $f\bar{\tau}[\psi]$ has the combined
conformal dimension $1/10+1/2=3/5$, and the third $s\bar{\tau}[\Sigma]$
has $3/80+9/16=3/5$. Although the super-sector splits into three
under $SO(9)_{1}$, it is irreducible under $(F_{4})_{1}$.
\end{document}